\documentclass[aps,pra,reprint,superscriptaddress,amsmath,amssymb,nofootinbib]{revtex4-2}
\usepackage{orcidlink}

\usepackage{graphicx}
\usepackage{dcolumn}
\usepackage{bm}

\usepackage{braket}
\usepackage{physics}
\usepackage{ulem}
\usepackage{soul}
\usepackage{tikz}
\usetikzlibrary{arrows.meta, positioning, shapes.geometric, calc}
\usepackage{xcolor}
 
\usepackage[T1]{fontenc}
\usepackage{hyperref}
\usepackage[utf8]{inputenc}

\DeclareUnicodeCharacter{2010}{-}    
\DeclareUnicodeCharacter{2011}{-}    
\DeclareUnicodeCharacter{2012}{-}    
\DeclareUnicodeCharacter{2013}{--}   
\DeclareUnicodeCharacter{2014}{---}  
\DeclareUnicodeCharacter{2212}{-}    

\DeclareUnicodeCharacter{2018}{`}    
\DeclareUnicodeCharacter{2019}{'}    
\DeclareUnicodeCharacter{201C}{``}   
\DeclareUnicodeCharacter{201D}{''}   

\newcommand{\Afull}{\mathcal{A}_{\mathrm{full}}}              
\newcommand{\Apart}[1]{\mathcal{A}_{#1}}                       
\newcommand{\LstarA}[1]{\mathcal{L}^{\star}_{\mathcal{A}_{#1}}}
\newcommand{\Lstarfull}{\mathcal{L}^{\star}_{\Afull}}

\begin{document}

\title{
Neural quantum states for entanglement depth certification from randomized Pauli measurements
}

\author{Marcin P\l{}odzie\'{n}\orcidlink{0000-0002-0835-1644}}
\email{marcin.plodzien@qilimanjaro.tech}
\affiliation{Qilimanjaro Quantum Tech, Carrer de Veneçuela 74, 08019 Barcelona, Spain}

\date{\today}

\begin{abstract}
Entanglement depth quantifies how many qubits share genuine multipartite entanglement, but certification typically relies on tailored witnesses or full tomography, both of which scale poorly with system size. We recast entanglement-depth and non-$k$-separability certification as likelihood-based model selection among neural quantum states whose architecture enforces a chosen entanglement constraint. A hierarchy of separable neural quantum states is trained on finite-shot local Pauli outcomes and compared against an unconstrained reference model trained on the same data. When all constrained models are statistically disfavored, the data certify entanglement beyond the imposed limit directly from measurement statistics, without reconstructing the density matrix. We validate the method on simulated six- and ten-qubit datasets targeting GHZ, Dicke, and Bell-pair states, and demonstrate robustness for mixed states under local noise. Finally, we discuss lightweight interpretability diagnostics derived from trained parameters that expose coarse entanglement patterns and qubit groupings directly from bitstring statistics.
\end{abstract}

\maketitle

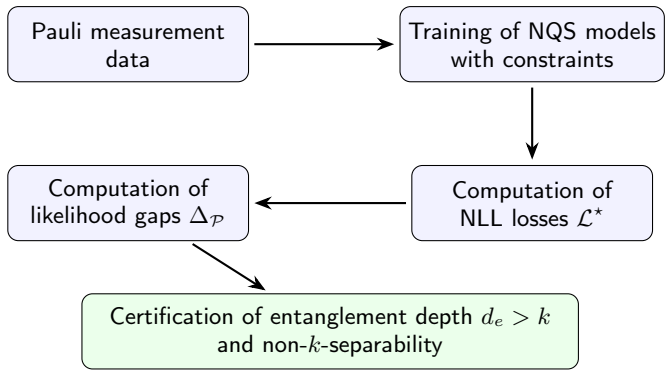
\begin{figure}[t!]
\centering
\begin{tikzpicture}[
    font=\sffamily,
    node distance=1.1cm and 2.0cm,
    box/.style={
        draw,
        rounded corners,
        align=center,
        minimum width=3.2cm,
        minimum height=1cm,
        fill=blue!5
    },
    arrow/.style={->, thick, >=Stealth, shorten >=2pt, shorten <=2pt}
]

\node[box] (data) {Pauli measurement\\data};
\node[box, right=of data] (train) {Training of NQS models\\with constraints};

\node[box, below=of data] (gaps) {Computation of\\likelihood gaps $\Delta_{\mathcal{P}}$};
\node[box, below=of train] (nll) {Computation of\\NLL losses $\mathcal{L}^\star$};

\node[box, below=1.2cm of $(gaps)!0.5!(nll)$, minimum width=6.6cm, fill=green!8]
(certify) {Certification of entanglement depth $d_e> {l}$\\and non-$k$-separability};

\draw[arrow] (data) -- (train);
\draw[arrow] (train) -- (nll);
\draw[arrow] (nll) -- (gaps);
\draw[arrow] (gaps) -- (certify);

\end{tikzpicture}
\caption{
\textbf{Protocol for entanglement-depth certification via generative modeling.}
Bitstrings obtained from randomized Pauli measurements on the target device are used to train a hierarchy of Neural Quantum State (NQS) models.
The hierarchy consists of an unconstrained reference model ($\mathcal{A}_{\text{full}}$) and various Separable NQS architectures ($\mathcal{A}_{\mathcal{P}}$), where the bipartite connectivity is masked to enforce factorization into clusters of size at most  {$l$}.
By minimizing the negative log-likelihood (NLL) $\mathcal{L}^*$, we compute the likelihood gap
$\Delta_{\mathcal{P}} = \mathcal{L}^*_{\mathcal{A}_{\mathcal{P}}} - \mathcal{L}^*_{\mathcal{A}_{\text{full}}}$.
A persistent gap ($\Delta_{\mathcal{P}} \gg 0$) signifies that the data cannot be explained by any state within the  {$l$}-producible hypothesis class, thereby certifying an entanglement depth $d_e >  {l}$ without reconstructing the full density matrix.
}
\label{fig:fig1}
\end{figure}

{
 
\section{Introduction}
Multipartite entanglement constitutes a fundamental resource for quantum information processing, metrology, and simulation. However, characterizing its structure in many-body systems remains a formidable experimental challenge. Beyond the simple dichotomy between separable and entangled states, a more granular measure is the \textit{entanglement depth}: the minimum number of particles that must be genuinely entangled to produce the state~\cite{Guhne2009Review,Sorensen2001SpinSqueezing,Toth2005TwoSettings,GuhneSeevinck2010,Bruss2016,Watrous2018,Wilde2017}. This concept is formally captured by the notion of  {$l$}-producibility, and closely related notions of $k$-separability used in multipartite entanglement classification; a state is  {$l$}-producible if it can be described---even probabilistically---as a mixture of pure states that factorize into clusters of size at most  {$l$}~\cite{Guhne2009Review,Huber2013Multidimensional}.

Conventional characterization strategies typically rely on either tailored entanglement witnesses or quantum tomography followed by post-processing~\cite{Guhne2009Review,GuhneSeevinck2010,Huber2013Multidimensional,Plodzien2024_non_k_sep}. Witness-based methods generally necessitate carefully engineered measurement settings and often rely on prior assumptions about the target state structure, such as graph-state or symmetric Dicke-state symmetries. Conversely, tomographic techniques suffer from a scaling problem: the number of required measurements and the classical post-processing costs grow exponentially with system size, rendering them intractable beyond a handful of qubits. Furthermore, modern quantum devices are often best treated as black boxes generating measurement outcomes in local bases. Ideally, one would extract entanglement information directly from these statistics without committing to a costly full reconstruction of the density matrix.

Machine learning (ML) offers a complementary avenue for data-driven assessment of entanglement properties \cite{ Goes2021, Hiesmayr2021,Krawczyk2024,Pawlowski2024,Taghadomi2025,Dawid2025_book}. A common supervised setting trains classifiers to distinguish separable from entangled states using tomographic or correlator-level features~\cite{Lu2018,Asif2023,Urea2024,Wieniak2023,Koutny2022,Palmieri2024,Bal2025}, while unsupervised variants aim to cluster entangled and separable data without labeled examples~\cite{Chen2021_QST}. Beyond binary detection, learned models have been employed to infer finer aspects of multipartite structure, including partition- or depth-related organization~\cite{Chen2021,Khoo2021,Li2024,Li2025}. In parallel, regression-based schemes estimate entanglement measures from experimentally accessible primitives, such as collective or weak measurements~\cite{Gray2018,Roik2022,Lin2023,Feng2024,Wang2025,Yang2024}. Related research replaces explicit reconstruction with learned generative models for basis-resolved outcomes, which can be queried for downstream state properties~\cite{Smith2021,Wei2024}, and kernel methods have also been explored for entanglement detection~\cite{Sabiote2025} (for review see~\cite{Varela2025,Jiao2024}).

In this work, we pursue a data-driven strategy distinct from supervised classification. Assuming access only to bitstrings from projective measurements in local Pauli bases, we investigate whether one can certify non-$k$-separability and bound the entanglement depth \textit{directly from how well different generative models explain these data}. Neural Quantum States (NQS) provide a natural framework for this task~\cite{Carleo2017,Deng2017,Melko2019,Medvidovi2024,Lange2024}. As highly expressive variational representations of many-body wave functions, NQS trained on measurement data serve as powerful generative models of the outcome distribution. This perspective has enabled neural-network approaches to quantum state tomography for pure states~\cite{Torlai2018NNQST,Beach2019QuCumber} and mixed states~\cite{Torlai2018LatentSpacePurification}, with successful demonstrations on experimental platforms~\cite{Torlai2019RydbergReconstruction,Neugebauer2020TwoQubitNNQST,Tiunov2020HomodyneML}.

We present a practical protocol for certifying non-$k$-separability and entanglement depth directly from finite-shot local Pauli measurement data, without making strong assumptions about the state. We train a fully entangling NQS and a hierarchy of  {separable neural quantum state (SNQS)} ansatzes---representing different multipartition structures---on the same bitstring data. The resulting hierarchy of optimized negative log-likelihoods serves as an operational witness of entanglement depth: likelihood gaps that persist for all architectures compatible with depth  {$l$} rule out  {$l$}-producibility, thereby certifying $d_e >  {l}$ without density-matrix reconstruction (see Fig.~\ref{fig:fig1}). Furthermore, we demonstrate that the parameters of the unrestricted NQS admit simple, interpretable post-processing. Effective visible--visible couplings and affinity matrices derived from the trained RBM weights yield informative signatures of the underlying structure---such as Bell-pair dimers or GHZ/Dicke clusters. We apply this methodology to three families of nontrivial multiqubit states, both pure and mixed, certifying entanglement depth in systems of up to ten qubits.

The paper is organized as follows.
In Sec.~\ref{sec:prelim} we review the notions of $k$-separability,
 {$l$}-producibility, and entanglement depth and introduce separable neural
quantum state architectures and their combinatorial structure.
In Sec.~\ref{sec:protocol} we describe the measurement protocol, the
negative log-likelihood training objective, and the likelihood-gap
criterion for non-$k$-separability.
Section~\ref{sec:results} presents numerical results for both pure and
mixed multi-qubit targets, including GHZ, Bell, and Dicke clusters under
local amplitude damping.
In Sec.~\ref{sec:interpretability} we analyze the trained unrestricted
NQS and introduce interpretable diagnostics that recover coarse
entanglement structure directly from the learned weights.
We conclude in Sec.~\ref{sec:conclusion}.
}

\section{Preliminaries}
\label{sec:prelim}

\subsection{Multipartite separability,  {$l$}-producibility, and entanglement depth}
\label{subsec:ksep_depth}

We consider an $N$-qubit system with Hilbert space
$\mathcal{H} = (\mathbb{C}^2)^{\otimes N}$.
Multipartite entanglement can be organized according to how the parties can
be partitioned into smaller groups, clusters, that share quantum
correlations.
The language of $k$-separability and  {$l$}-producibility is standard in the entanglement-depth literature
\cite{Wilde2017, Bruss2016, Watrous2018}.

Let $\{1,\ldots,N\}$ denote the set of qubit labels and
$\mathcal{P}=\{C_1,\ldots,C_k\}$ a partition into $k$ non-empty, disjoint subsets $C_b$ with union $\{1,\ldots,N\}$.
A pure state $\ket{\Psi}\in\mathcal{H}$ is called
\textit{$k$-separable with respect to $\mathcal{P}$} if it factorizes as
\begin{equation}
  \ket{\Psi}
  =
  \bigotimes_{b=1}^{k} \ket{\phi_{C_b}},
  \qquad
  \ket{\phi_{C_b}} \in
  \bigotimes_{i\in C_b} \mathbb{C}^2 .
\end{equation}
If there exists at least one partition $\mathcal{P}$ with $k\ge2$ such that
this holds, we say that $\ket{\Psi}$ is \textit{$k$-separable}.
The extreme cases are
$k=N$, where every cluster contains a single qubit and the state is fully
separable, and $k=1$, where the whole register forms one cluster and the
state may be genuinely $N$-partite entangled.
A pure state that is not $k$-separable for any $k\ge2$ is called
\textit{genuinely multipartite entangled}
 {(note that $k=1$ means no partition with two or more clusters exists under which the state factorizes)}
\cite{Guhne2009Review,GuhneSeevinck2010}.

For mixed states the definition is extended by convex combinations.
A density operator $\rho$ on $\mathcal{H}$ is called \textit{$k$-separable}
if it admits a decomposition
\begin{equation}
  \rho
  =
  \sum_{\lambda} p(\lambda)\,
  \bigotimes_{b=1}^{k}
     \rho^{(b)}_{C_b}(\lambda),
  \qquad
  p(\lambda)\ge0,\ \sum_\lambda p(\lambda)=1,
  \label{eq:ksep_mixed}
\end{equation}
where for each label $\lambda$ the operators
$\rho^{(b)}_{C_b}(\lambda)$ act on the clusters $C_b$ of some
partition $\mathcal{P}(\lambda)$ of $\{1,\ldots,N\}$ into $k$ groups.
Importantly, the partition $\mathcal{P}(\lambda)$ is in general allowed to
depend on $\lambda$; Eq.~\eqref{eq:ksep_mixed} thus describes classical
mixtures of possibly different $k$-separable pure states.
States that cannot be written in the form~\eqref{eq:ksep_mixed} are called
\textit{non-$k$-separable}.  In particular, non-biseparable states
($k=2$) are genuinely multipartite entangled
\cite{Guhne2009Review,GuhneSeevinck2010,Huber2013Multidimensional}.

While $k$-separability counts the number of entangled blocks, it is often
more natural to characterize \textit{how many} qubits are entangled within a
block.  This leads to the notion of  {$l$}-producibility and entanglement
depth, originally introduced in the context of spin-squeezed and
many-body states
\cite{Sorensen2001SpinSqueezing,Toth2005TwoSettings,Guhne2009Review}.
 {To avoid notational ambiguity we use distinct symbols in the two definitions: $k$ denotes the number of clusters in a $k$-separable partition, while $l$ bounds the maximal cluster size in $l$-producibility. In much of the literature both roles are assigned to the same letter; our convention follows the usage of Refs.~\cite{Guhne2009Review,GuhneSeevinck2010,Wilde2017} while eliminating the resulting notational clash.}

A pure state $\ket{\Psi}$ is called \textit{ {$l$}-producible} if it can be
written as a product of clusters containing at most  {$l$} qubits each,
\begin{equation}
  \ket{\Psi}
  =
  \bigotimes_{b=1}^{m}
    \ket{\phi_{C_b}}, 
   \qquad
   |C_b| \le  {l},\quad \bigcup_{b} C_b = \{1,\ldots,N\}.
\end{equation}
For mixed states, $\rho$ is  {$l$}-producible if it is a convex combination of
 {$l$}-producible pure states,
\begin{equation}
  \rho
  =
  \sum_{\lambda} p(\lambda)\,
   \ket{\Psi_\lambda}\bra{\Psi_\lambda},
  \quad
  \ket{\Psi_\lambda} \text{  {$l$}-producible for every }\lambda.
  \label{eq:kproducible}
\end{equation}
Intuitively,  {$l$}-producible states can be prepared by entangling at most
 {$l$} parties at a time.  Any state that is not  {$l$}-producible is said to be
\textit{non- {$l$}-producible} and has entanglement extending over at least
 {$l+1$} qubits.

The \textit{entanglement depth} $d_e(\rho)$ is then defined as the smallest
integer  {$l$} such that $\rho$ is  {$l$}-producible,
\begin{equation}
  d_e(\rho)
  =
  \min\{\, {l} \;|\; \rho \text{ is  {$l$}-producible}\,\}.
\end{equation}
Fully separable states have $d_e=1$; states with $d_e=2$ contain only
pairwise entanglement (e.g.\ products of Bell pairs), and genuinely
multipartite entangled states on $N$ qubits have $d_e = N$ and are
non-$(N-1)$-producible.

In this work we exploit the correspondence between such
partition structures and the connectivity masks of separable neural-network
states (SNQS).
Each SNQS architecture implements a specific partition
$\mathcal{P}=\{C_1,\dots,C_B\}$ by constraining the visible--hidden
connectivity of the restricted Boltzmann machine so that hidden units
couple only to qubits within the same cluster.
The maximal cluster size $|C_b|$ in a given architecture therefore
bounds the entanglement depth of all states representable by that ansatz.
By training both an unconstrained NQS and families of SNQS with different
clusterings on the same measurement data and comparing their optimal
negative log-likelihoods, we obtain an operational tool to falsify
 {$l$}-producibility hypotheses and to certify non-$k$-separability and
entanglement depth \textit{directly from finite-shot statistics}.


\subsection{Neural quantum states and separable architectures}
\label{sec:nqs}

\begin{figure}[t]
  \centering
  \includegraphics[width=\columnwidth]{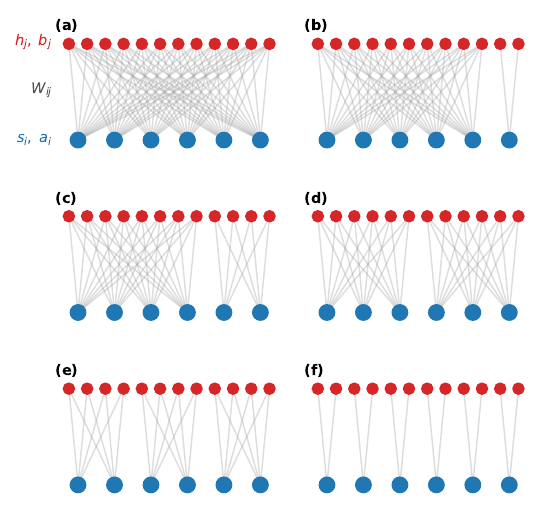}
  \caption{{Restricted Boltzmann machine connectivity for $N=6$ visible units $s_i$ (blue, bottom row) with biases $a_i$, and $M=12$ hidden units $h_j$ (red, top row) with biases $b_j$, connected by the weight matrix $W_{ij}$ (grey lines).
  Each panel corresponds to a different partition $\mathcal{P}$ that defines a separable neural quantum state (SNQS) architecture:
  (a)~unconstrained RBM ($\Afull$, $d_{\max}=6$), where every visible unit connects to every hidden unit;
  (b)~$\Apart{5|1}$ ($d_{\max}=5$);
  (c)~$\Apart{4|2}$ ($d_{\max}=4$);
  (d)~$\Apart{3|3}$ ($d_{\max}=3$);
  (e)~$\Apart{2|2|2}$ ($d_{\max}=2$);
  (f)~$\Apart{1|1|1|1|1|1}$ ($d_{\max}=1$, fully separable product ansatz).
  The absence of inter-block connections (visible as spatial gaps) enforces the block-product structure of Eq.~(9), bounding the entanglement depth of any representable state by $d_{\max}(\mathcal{P})$.}}
  \label{fig:rbm_architectures}
\end{figure}

Neural quantum states (NQS) provide a variational parametrization of
many-body wave functions in terms of artificial neural networks
\cite{Carleo2017,Melko2019,Torlai2018,Medvidovi2024,Lange2024}.
 {A standard architecture for this purpose is the restricted Boltzmann machine (RBM)~\cite{Carleo2017}. An RBM is a bipartite undirected graphical model with two layers of binary variables: a \emph{visible} layer encoding the physical spin configuration $\mathbf{s}\in\{\pm 1\}^N$, and a \emph{hidden} layer of auxiliary units $\mathbf{h}\in\{\pm 1\}^M$ that induce effective many-body correlations among the visible spins. Because connections exist exclusively \emph{between} layers (via a weight matrix $W_{ij}$) with no intra-layer couplings (Fig.~\ref{fig:rbm_architectures}(a)), the hidden variables can be traced out analytically. This yields a tractable closed-form expression for the unnormalized amplitude (Eq.~\eqref{eq:rbm_wavefunction}), meaning the Born-rule probabilities $p_\theta(\mathbf{s}) = |\Psi_\theta(\mathbf{s})|^2$ can be evaluated exactly for moderate system sizes.}
 
We consider $N$ qubits with computational basis
$\{\ket{\mathbf{s}}\}_{\mathbf{s}\in\{\pm1\}^N}$.
An RBM consists of $N$ visible spins $s_i\in\{\pm1\}$ and $M$ hidden spins
$h_j\in\{\pm1\}$ with energy
\begin{equation}
  E_\theta(\mathbf{s},\mathbf{h})
  = -\sum_{i} a_i s_i
    -\sum_{j} b_j h_j
    -\sum_{ij} s_i W_{ij} h_j,
\end{equation}
where $\theta=\{a_i,b_j,W_{ij}\}$ are (complex) parameters.
Tracing out the hidden layer yields an unnormalized amplitude
\begin{equation}\label{eq:rbm_wavefunction}
\begin{split}
  \tilde{\Psi}_\theta(\mathbf{s})
  &= \sum_{\mathbf{h}}
      \exp\bigl[-E_\theta(\mathbf{s},\mathbf{h})\bigr]\\
  &= \exp\Bigl(\sum_i a_i s_i\Bigr)
    \prod_j 2\cosh\Bigl(b_j+\sum_i W_{ij}s_i\Bigr),
    \end{split}
\end{equation}
and the normalized neural quantum state is
\begin{equation}
  \ket{\Psi_\theta}
  = \frac{1}{\sqrt{\mathcal{N}_\theta}}
    \sum_{\mathbf{s}}
      \tilde{\Psi}_\theta(\mathbf{s}) \ket{\mathbf{s}},
  \qquad
  \mathcal{N}_\theta
  = \sum_{\mathbf{s}} |\tilde{\Psi}_\theta(\mathbf{s})|^2.
\end{equation}
In practice, amplitude and phase are modeled via two real
RBMs \cite{Carleo2017,Torlai2018},
$\Psi_\theta(\mathbf{s}) = \exp[A_\theta(\mathbf{s})
+ i \Phi_\theta(\mathbf{s})]$.
The Born probabilities
$p_\theta(\mathbf{s}) = |\Psi_\theta(\mathbf{s})|^2$
define a generative model for measurement outcomes in the computational
basis.  

The expressive power of the RBM can be constrained by masking the
visible--hidden connectivity according to a chosen partition of the
qubits.
Let $\mathcal{P}=\{C_1,\dots,C_B\}$ be a partition of
$\{1,\dots,N\}$ into disjoint clusters.
We divide the hidden units into disjoint groups $H_1,\dots,H_B$ and
require that units in $H_b$ may only connect to visible qubits in $C_b$
\cite{Harney2020NJP,Harney2021Mixed}.
All weights $W_{ij}$ violating this rule are set to zero
 {(Fig.~\ref{fig:rbm_architectures}(b)--(f))}.

With this mask the energy decomposes into a sum of cluster energies,
$E_\theta(\mathbf{s},\mathbf{h})
 = \sum_{b=1}^{B} E^{(b)}(\mathbf{s}_{C_b},\mathbf{h}_{H_b}),
$ and the unnormalized amplitude factorizes,
\begin{equation}
  \tilde{\Psi}_\theta(\mathbf{s})
  = \prod_{b=1}^{B}
       {\tilde{\psi}_b}(\mathbf{s}_{C_b}),
\end{equation}
so that the normalized neural quantum state has the form
\begin{equation}
\ket{\Psi_\theta}
  = \bigotimes_{b=1}^{B} \ket{\phi_b},
\end{equation}
with each factor $\ket{\phi_b}$ supported on the cluster $C_b$.
Thus every SNQS implementing the partition $\mathcal{P}$ represents an
$B$-separable state, with
maximal cluster size
$d_{\max}(\mathcal{P}) = \max_b |C_b|$.
The entanglement depth of any state realizable by this architecture is
therefore bounded by $d_{\max}(\mathcal{P})$.

Decreasing $d_{\max}(\mathcal{P})$ yields a hierarchy of increasingly
constrained SNQS families:
from fully separable product states ($|C_b|=1$ for all $b$),
through bipartite and tripartite clusterings,
up to the unconstrained RBM where
$\mathcal{P}=\{\{1,\dots,N\}\}$ and $d_{\max}(\mathcal{P})=N$.
This architectural hierarchy is the backbone of our non-$k$-separability
certification protocol.

In the following we denote by $\Apart{\mathcal{P}}$ the SNQS family
implementing partition $\mathcal{P}$.
For an $N$-qubit system we often use an explicit label
$\mathcal{P} = n_1|n_2|\dots|n_B$ describing the block sizes
in some fixed ordering of qubits.
For instance, for $N=6$ we write
\begin{equation}
  \Apart{2|2|2}
\end{equation}
for the three-block SNQS where qubits are grouped as
$\{0,1\}|\{2,3\}|\{4,5\}$, and
\begin{equation}
  \Apart{1|1|1|1|1|1}
\end{equation}
for the fully separable product ansatz.
The corresponding optimized losses are written as
$\LstarA{2|2|2}$ and $\LstarA{1|1|1|1|1|1}$, respectively.
The fully connected reference NQS is denoted by $\Afull$ with optimal
loss $\Lstarfull$.

The SNQS hierarchy relies on selecting a partition
$\mathcal{P} = \{C_{1},\dots,C_{B}\}$ of the qubit index set
$\{1,\dots,N\}$ into nonempty, disjoint subsets (blocks).
From the viewpoint of combinatorics, this is precisely the problem of
counting set partitions, a classical topic connected to the Stirling
numbers of the second kind and the Bell numbers
\cite{comtet1974,stanley2012,graham1994concrete,cameron1994combinatorics}.

The number of ways to partition an $N$--element set into \textit{exactly}
$k$ nonempty, unlabeled subsets is given by a Stirling number of the
second kind,
\begin{equation}
  S(N,k)
  = \frac{1}{k!}\sum_{i=0}^{k} (-1)^{k-i}
      \binom{k}{i} \, i^{N},
  \label{eq:stirling-second-kind}
\end{equation}
which counts the number of distinct equivalence relations on
$\{1,\dots,N\}$ with exactly $k$ equivalence classes.
These numbers satisfy the recurrence
\begin{equation}
  S(N{+}1,k)
  = k\,S(N,k) + S(N,k{-}1),
  \qquad
  S(0,0)=1,
\end{equation}
with $S(N,0)=0$ for $N>0$, and have the exponential generating function
\begin{equation}
  \sum_{N=0}^{\infty} S(N,k)\,\frac{z^{N}}{N!}
  = \frac{(e^{z}-1)^{k}}{k!}.
\end{equation}

Summing over all $k$ yields the total number of multipartitions of an
$N$--element set,
\begin{equation}
  B_{N}
  = \sum_{k=0}^{N} S(N,k),
  \label{eq:Bell-number}
\end{equation}
known as the $N$th \textit{Bell number}.  The Bell numbers obey
\begin{equation}
  B_{N+1}
  = \sum_{k=0}^{N} \binom{N}{k} B_{k},
\end{equation}
and admit the exponential generating function
\begin{equation}
  \sum_{N=0}^{\infty} B_{N}\,\frac{z^{N}}{N!}
  = e^{e^{z}-1}.
\end{equation}

Already for $N=6$ the Stirling numbers take the values
$S(6,1)=1$, $S(6,2)=31$, $S(6,3)=90$, $S(6,4)=65$, $S(6,5)=15$,
and $S(6,6)=1$, so that $B_{6}=203$ distinct multipartitions exist.
These include, for example, the fully separable partition
$1|1|1|1|1|1$, the ``dimerized'' structure $2|2|2$, and the bipartition
$3|3$.
The combinatorial explosion of $B_{N}$ with $N$
($B_{8}=4140$, $B_{10}=115\,975$, \textit{etc.}) shows that for even
moderate system sizes an \textit{exhaustive} scan over all partitions is
neither necessary nor practical.
In the SNQS setting one therefore designs a \textit{hierarchy} of
architectures based on a carefully chosen subset of multipartitions
(e.g., by increasing maximal block size or by exploiting physical
symmetries), which is sufficient to probe the non-$k$-separability
structure relevant for entanglement-depth certification.

\section{Measurement protocol and likelihood-based certification}
\label{sec:protocol}

On a real device the quantum state is not directly accessible; only measurement statistics in local bases are available. We therefore model the unknown target as a density operator $\rho_{\mathrm{tgt}}$ that can be probed only through a finite collection of measurement outcomes, summarized by empirical frequencies $f_b(\mathbf{s})$ in local measurement bases. For a given hierarchical family of ansatz states, we quantify how well each level of the hierarchy can reproduce these observed frequencies by minimizing the negative log-likelihood. The resulting likelihood values, and in particular the gaps between successive levels in the hierarchy, indicate how much expressive power is required to explain the data. Our objective is to use these likelihood gaps to certify non-$k$-separability and entanglement depth directly from the measurement statistics, without reconstructing $\rho_{\mathrm{tgt}}$ tomographically.

We consider projective measurements in local single-qubit Pauli bases.
A measurement setting is specified by a basis label
$b = (b_1,\dots,b_N)$ with $b_i \in \{X,Y,Z\}$ indicating whether qubit
$i$ is measured in the $X$, $Y$, or $Z$ eigenbasis.
In a given shot we choose a basis pattern $b$, prepare the state
$\rho_{\mathrm{tgt}}$, measure all qubits in that basis, and record the
classical bitstring
$\mathbf{s} = (s_1,\dots,s_N) \in \{0,1\}^N$.

Collecting $N_{\mathrm{shots}}$ repetitions over possibly many
different bases yields a dataset
\begin{equation}
  \mathcal{D}
  = \{(b_\ell,\mathbf{s}_\ell)\}_{\ell=1}^{N_{\mathrm{shots}}}.
  \label{eq:data-def}
\end{equation}
For each basis $b$ and outcome $\mathbf{s}$ we define empirical
frequencies
\begin{equation}
  f_b(\mathbf{s})
  = \frac{1}{N_b}
    \sum_{\ell:\,b_\ell = b}
    \delta_{\mathbf{s},\mathbf{s}_\ell},
\end{equation}
where $N_b$ is the total number of shots taken in basis $b$.
In the limit $N_b\to\infty$ these converge to the Born probabilities
\begin{equation}
  p_{\mathrm{tgt}}(\mathbf{s}\mid b)
  = \Tr\!\left[\rho_{\mathrm{tgt}}
               \,\Pi^{(b)}_{\mathbf{s}}\right],
\end{equation}
with $\Pi^{(b)}_{\mathbf{s}}$ the projector onto the outcome
$\mathbf{s}$ in basis $b$.

In the numerical experiments we mimic this protocol by generating bitstrings from ideal target
states, in a set of local Pauli bases $(X,Y,Z)$ with randomly chosen
patterns $b$.

Given a neural ansatz architecture $\mathcal{A}$ with parameters
$\theta$, the corresponding model state defines a density
operator $\rho_\theta$.
For each basis $b$ and outcome $\mathbf{s}$ the model predicts Born
probabilities
\begin{equation}
  p_{\theta}(\mathbf{s}\mid b)
  = \Tr\!\left[\rho_{\theta}\,
               \Pi^{(b)}_{\mathbf{s}}\right].
  \label{eq:model-probs}
\end{equation}

The natural statistical objective is the (average) negative
log-likelihood (NLL) of the observed dataset $\mathcal{D}$:
\begin{equation}
  \mathcal{L}(\mathcal{A};\theta)
  = - \frac{1}{N_{\mathrm{shots}}}
      \sum_{\ell=1}^{N_{\mathrm{shots}}}
      \log p_{\theta}(\mathbf{s}_\ell \mid b_\ell).
  \label{eq:nll-single}
\end{equation}
Grouping identical outcomes into frequencies yields the equivalent form
\begin{equation}
  \mathcal{L}(\mathcal{A};\theta)
  = - \sum_{b} \sum_{\mathbf{s}}
      f_b(\mathbf{s})
      \log p_{\theta}(\mathbf{s}\mid b).
  \label{eq:nll-total}
\end{equation}
Minimizing~\eqref{eq:nll-total}  drives the NQS to reproduce
the Born probabilities of the unknown target state as closely as the
model family allows.

\subsection{Likelihood gaps and non-\texorpdfstring{$k$}{k}-separability}
\label{subsec:nll-gaps}

We index each SNQS architecture by the partition
$\mathcal{P} = \{C_1,\dots,C_M\}$ of the qubits that it implements.
We write $\mathcal{A}_{\mathcal{P}}$ for the corresponding ansatz
family and $\mathcal{A}_{\mathrm{full}}$ for the unconstrained (fully
connected) NQS.
Given a fixed measurement dataset and an architecture $\mathcal{A}$ with
parameters $\theta$, the NLL is $\mathcal{L}(\mathcal{A};\theta)$ as in
Eq.~\eqref{eq:nll-total}.  Let $\mathcal{L}^\star_{\mathcal{A}}$ denote
the minimal NLL achieved by architecture $\mathcal{A}$ after training,
\begin{equation}
  \mathcal{L}^\star_{\mathcal{A}}
  = \min_{\theta} \mathcal{L}(\mathcal{A};\theta).
\end{equation}
For a partition $\mathcal{P}$ we then define the \textit{loss gap}
\begin{equation}
  \Delta_{\mathcal{P}}
  = \mathcal{L}^\star_{\mathcal{A}_{\mathcal{P}}}
    - \mathcal{L}^\star_{\mathcal{A}_{\mathrm{full}}} .
  \label{eq:def-loss-gap}
\end{equation}
 {At the level of exact global optima, $\Delta_{\mathcal{P}} \ge 0$ by model-class inclusion: the SNQS class is a strict subset of the unconstrained NQS class (obtained by zeroing specific weights), so the globally optimal NLL of the unconstrained model is guaranteed to be no worse.}
$\Delta_{\mathrm{full}} = 0$, where ``full'' denotes
the partition with a single block containing all qubits.
 {In finite optimization runs, small violations of this bound can occur; we address this through the heuristic threshold $\overline{\Delta}_{\mathrm{th}}$ introduced below.}

Intuitively, $\Delta_{\mathcal{P}}$ measures how much worse the best
$\mathcal{P}$-separable ansatz performs compared to the best
unconstrained ansatz when both are fit to the same experimental
bitstring histograms.  In the ideal limit of tomographically
informational data, sufficiently expressive NQS ansatzes, and perfect
optimization, the quantities $\mathcal{L}^\star_{\mathcal{A}}$ converge
to the minimal cross-entropy between the data distribution and the model
family $\mathcal{A}$.  If the true state $\rho_{\mathrm{tgt}}$ were
exactly representable within $\mathcal{A}_{\mathcal{P}}$, then one would
obtain $\Delta_{\mathcal{P}}\approx 0$ up to finite-shot fluctuations.

For each partition $\mathcal{P}$ the associated constrained NQS ansatz
$\mathcal{A}_{\mathcal{P}}$ only represents states whose entanglement
depth is bounded by
\begin{equation}
  d_{\max}(\mathcal{P})
  = \max_{C\in\mathcal{P}} |C|.
\end{equation}

{
If, for a fixed integer  {$l$}, we find that
\begin{equation}
\Delta_{\mathcal{P}} > \overline{\Delta}_{th} \quad \forall \mathcal{P} \text{ such that } d_{max}(\mathcal{P}) \le  {l},
\label{eq:gap-kprod}
\end{equation}
then the experimental data are incompatible with all  {$l$}-producible states represented in our constrained NQS hierarchy. In this proof-of-principle study, we determine $\overline{\Delta}_{th}$ heuristically based on the typical stability of the loss function after convergence. We treat likelihood differences comparable to the residual optimization fluctuations (typically $\lesssim 10^{-3}$ in our numerical experiments) as statistically insignificant, whereas structurally excluded partitions exhibit gaps orders of magnitude larger.

 {In this situation, we certify that $\rho_{tgt}$ is not $l$-producible and obtain a lower bound on the entanglement depth,}
}
\begin{equation}\label{eq:NLL_ent_criterion}
  d_e(\rho_{\rm tgt}) >  {l} .
\end{equation}
In the numerical examples below we explicitly construct such hierarchies
of SNQS architectures, compute the optimized losses
$\mathcal{L}^\star_{\mathcal{A}_{\mathcal{P}}}$, and report the loss
gaps $\Delta_{\mathcal{P}}$.
 {For conciseness, we summarize the resulting likelihood orderings across partitions using the compact notation}
\begin{equation}
  \mathcal{L}^\star_{\mathcal{A}_{3|3}} \approx
  \mathcal{L}^\star_{\mathcal{A}_{\mathrm{full}}},
  \qquad
  \mathcal{L}^\star_{\mathcal{A}_{2|2|2}} \gg
  \mathcal{L}^\star_{\mathcal{A}_{\mathrm{full}}},
\end{equation}
 {where the illustrative $3|3$ and $2|2|2$ partitions refer to a six-qubit system.}

 {\subsection{Scope and limitations}
\label{subsec:practical}}

 {The total number of set partitions of $N$ qubits (the Bell number $B_N$) grows super-exponentially. In practice, however, an exhaustive scan over all architectures is neither required nor intended.
When characterizing near-term devices, the hardware layout and the executed circuit dictate the relevant cuts: the user typically tests whether entanglement survives noise across specific chip boundaries. For such targeted tests, the computational cost is independent of $B_N$.
Each training run scales as $O(E \cdot 2^N \cdot N \cdot H)$, where $E$ is the number of epochs, $N$ the number of qubits, and $H$ the hidden-layer size; for the present proof-of-principle study ($N \leq 10$) this is tractable. For larger systems, sampling-based NLL estimators---as routinely used in variational Monte Carlo~\cite{Carleo2017}---can replace exact enumeration, and more expressive autoregressive neural quantum states~\cite{Sharir2020} can serve as drop-in replacements for RBMs, with partition constraints enforced by masking the autoregressive conditionals.}

 {Our experiments use 200 randomly chosen local Pauli basis strings, each sampled with 2000 shots. The total budget of $4\times 10^5$ outcomes is comparable to that of randomized measurement protocols like classical shadows. Because the NLL objective~\eqref{eq:nll-total} is a per-shot average, the leading statistical fluctuations decrease as $1/\sqrt{N_{\mathrm{shots}}}$, though the prefactor may depend on system size, the gap magnitude, and the observable structure.}

 {A large likelihood gap $\Delta_{\mathcal{P}}\gg 0$ should be interpreted as a one-sided, variational exclusion of the corresponding constrained hypothesis class: it provides evidence that the measured data are incompatible with that $l$-producible model family. By contrast, a small gap is inconclusive, as it may result either from true compatibility or from limited expressivity / imperfect optimization of the unconstrained ansatz. In practice, this interpretation relies on sufficiently good convergence of both constrained and unconstrained trainings; poor optimization of the constrained model can artificially enlarge the gap. For this reason, throughout the paper we treat the certification as variational with respect to the chosen model class and optimization procedure, and we use the Hilbert--Schmidt fidelity diagnostic introduced below as a sanity check on training quality. Additionally, because the unconstrained model has strictly more parameters, small likelihood improvements may partly reflect capacity to fit finite-shot noise rather than genuine entanglement; for larger systems, complexity-penalized model selection or held-out validation can sharpen this distinction. Finally, while we demonstrate the hierarchy using RBMs, the framework accommodates any generative model family with partition-constrained variants; using more expressive architectures tightens the certification lower bounds.}

 {To confirm that the NLL minimum corresponds to an accurate generative model, we employ the following diagnostic.}
Whenever the exact target state $\rho_{\mathrm{tgt}}$ is available, as
in the synthetic benchmarks of Sec.~\ref{sec:results}, we additionally
compare it with the reconstructed state
$\rho_{\mathrm{est}}^{(\mathcal{A}_{\mathcal{P}})}$ obtained from a
given ansatz $\mathcal{A}_{\mathcal{P}}$ using the
{Hilbert--Schmidt (HS) overlap}
\begin{equation}
  F_{\mathrm{HS}}\!\left(\rho_{\mathrm{tgt}},
                         \rho_{\mathrm{est}}^{(\mathcal{A}_{\mathcal{P}})}\right)
  =
 \frac{\Tr\!\bigl[\rho_{\mathrm{tgt}}
                   \rho_{\mathrm{est}}^{(\mathcal{A}_{\mathcal{P}})}\bigr]}{\sqrt{\Tr[\rho_{\rm tgt}^2] \Tr[{\rho_{\mathrm{est}}^{(\mathcal{A}_{\mathcal{P}})}}^2]}}
  \label{eq:hs-fid-main}
\end{equation}
and the associated HS distance
\begin{equation}
  D_{\mathrm{HS}}\!\left(\rho_{\mathrm{tgt}},
                         \rho_{\mathrm{est}}^{(\mathcal{A}_{\mathcal{P}})}\right)
  =
  \left\|
    \rho_{\mathrm{tgt}}
    - \rho_{\mathrm{est}}^{(\mathcal{A}_{\mathcal{P}})}
  \right\|_2
  = \sqrt{\Tr\!\Bigl[
        \bigl(\rho_{\mathrm{tgt}}
              - \rho_{\mathrm{est}}^{(\mathcal{A}_{\mathcal{P}})}\bigr)^{2}
      \Bigr]}.
  \label{eq:hs-dist-main}
\end{equation}
Both quantities are analytic and easy to evaluate numerically
\cite{Miszczak2009,NielsenChuang2010}.  Importantly,
$F_{\mathrm{HS}}$ and $D_{\mathrm{HS}}$ are used here only as
\textit{diagnostics} in numerical experiments; the actual
non-$k$-separability and entanglement-depth certification on
experimental data relies solely on the likelihood gaps
$\Delta_{\mathcal{P}}$ obtained from finite-shot statistics, without
any full tomographic reconstruction of $\rho_{\mathrm{tgt}}$.

{
 
\section{Numerical results}
\label{sec:results}

In this section we illustrate the likelihood--gap certification strategy
on three synthetic pure $N$-qubit targets of increasing structural
complexity: a six-qubit GHZ state, a product of three Bell pairs, and a
ten-qubit product of heterogeneous entangled clusters.
In all cases we generate finite-shot data from the target state by
measuring in a pool of $200$ random local Pauli bases (with $2000$ shots
per basis), and we train both the unrestricted NQS ansatz $\Afull$ and a
family of SNQS ansatzes $\mathcal{A}_{\mathcal{P}}$ corresponding to
different multipartitions $\mathcal{P}$ of the qubits.

For each architecture we monitor the negative log--likelihood (NLL)
during training and record the minimal value $\LstarA{\mathcal{P}}$
attained along the optimization trajectory.  
The likelihood gap
$\Delta_{\mathcal{P}} = \LstarA{\mathcal{P}} - \Lstarfull$, defined in
Eq.~\eqref{eq:def-loss-gap}, is then used to certify non-$k$-separability
and entanglement depth according to the criterion
\eqref{eq:gap-kprod}.
Whenever the exact target state is known (as here), we additionally
evaluate Hilbert--Schmidt (HS) fidelity $F_{\mathrm{HS}}$ and the
associated HS distance $D_{\mathrm{HS}}$ as diagnostics; however, the
certification itself is phrased purely in terms of likelihood gaps and
does not rely on access to $\rho_{\mathrm{tgt}}$. For numerical details, see App.\ref{app:numerical_details}.

\begin{figure}
    \includegraphics[width=0.99\linewidth]{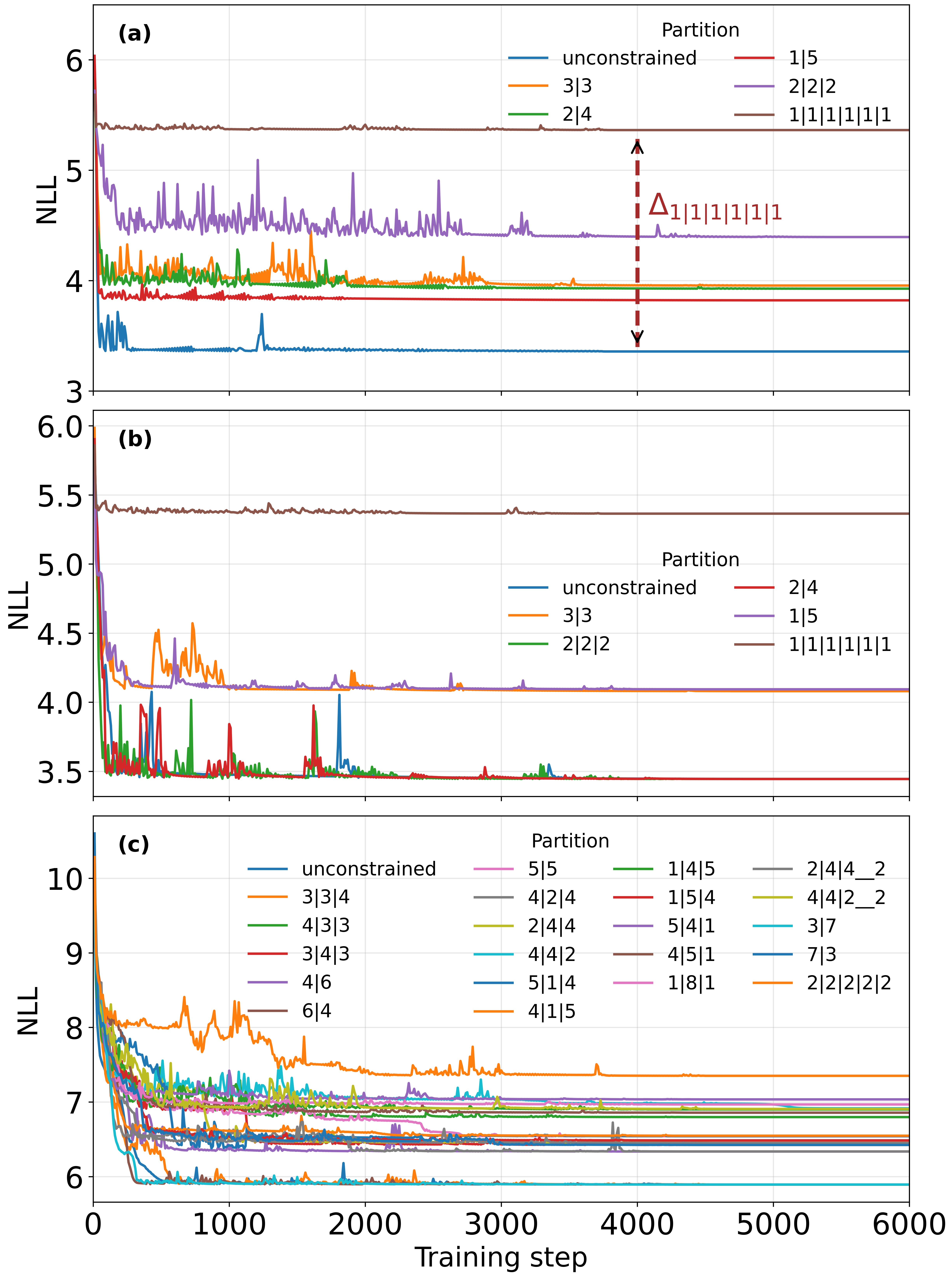}
\caption{
Training dynamics of the negative log--likelihood (NLL) for neural quantum
state models fitted to finite--shot random--Pauli data. In each panel, the
blue curve shows the fully entangling (``unconstrained'') ansatz, while the
other curves correspond to separable neural quantum states with different
partitions \(\mathcal{P}\) of the qubits (see legends), whose maximal block
size bounds the entanglement depth. 
(a) Six--qubit GHZ target, Eq.\eqref{eq:psi_A}.
(b) Product of three Bell pairs on six qubits, Eq.\eqref{eq:psi_B}.
(c) Ten--qubit product of a three--qubit GHZ state, a three--qubit Dicke state
with one excitation, and a four--qubit GHZ state in local bases, Eq.\eqref{eq:psi_C}.
Persistent offsets between the constrained and unconstrained NLL traces signal
partitions that are incompatible with the target entanglement structure. The NLL gap for partition $1|1|1|1|1|1$, $\Delta_{1|1|1|1|1|1}$ is denoted as dashed-brown vertical line.
}
\label{fig:fig2}
\end{figure}

Figure~\ref{fig:fig2} shows the training dynamics of the NLL
for a representative subset of partitions in each example, while the
optimized losses and fidelities are summarized in the tables below. 

\subsection{Pure states}
\subsubsection{Six-qubit GHZ}

Our first benchmark is the pure six-qubit GHZ state in the
computational basis,
\begin{equation}
  \label{eq:psi_A}
  \ket{\psi_{\mathrm{A}}}
  \equiv
  \ket{\psi_{\mathrm{GHZ}}^{\mathrm{ZZZZZZ}}}
  =
  \frac{1}{\sqrt{2}}\bigl(\ket{0_z}^{\otimes 6} + \ket{1_z}^{\otimes 6}\bigr),
\end{equation}
which is genuinely multipartite with entanglement depth
$d_e(\ket{\psi_{\mathrm{A}}}) = 6$.
We train the global (unrestricted) ansatz $\Afull$ together with SNQS
ansatzes associated with the partitions
\[
  \mathcal{P} \in
  \{3|3,\; 2|4,\; 1|5,\; 2|2|2,\; 1|1|1|1|1|1\},
\]
whose maximal block sizes are
$d_{\max}(\mathcal{P}) \in \{3,4,5,2,1\}$.

The global NQS converges to the lowest NLL, which we take as the
reference $\Lstarfull$.
All SNQS models exhibit sizeable positive loss gaps,
\[
  \Delta_{1|5} \approx 0.46,\qquad
  \Delta_{2|4} \approx 0.57,\qquad
  \Delta_{3|3} \approx 0.60,
\]
while more strongly constrained partitions lead to even larger gaps,
\[
  \Delta_{2|2|2} \approx 1.04,\qquad
  \Delta_{1|1|1|1|1|1} \approx 2.01.
\]
Thus every partition with $d_{\max}(\mathcal{P}) \le 5$ incurs a clear,
statistically significant likelihood penalty relative to the
unrestricted model (Table~\ref{tab:A_pure}).

Within our SNQS hierarchy, the measurement data are therefore
incompatible with all  {$l$}-producible models with  {$l \le 5$}, and the
likelihood--gap criterion, Eq.~\eqref{eq:NLL_ent_criterion}, certifies
$d_e(\rho_{\mathrm{tgt}}) > 5$,
in agreement with the known entanglement depth of the GHZ state.
This illustrates a pronounced ``entanglement barrier'': any attempt to
describe $\ket{\psi_{\mathrm{GHZ}}^{\mathrm{ZZZZZZ}}}$ with strictly
bounded entanglement depth leads to a robust increase in NLL.
For the global ansatz the reconstructed state achieves
$F_{\mathrm{HS}} \approx 1$ and very small HS distance, confirming that
the minimum of the NLL landscape corresponds to an accurate model of
the target and that the observed gaps are not optimization artefacts.

\subsubsection{Three Bell pairs}

\begin{table}[t]
  \centering
  \begin{tabular}{|c|c|c|c|c|}
    \hline\hline
    Partition $\mathcal{P}$ &  {$d_{\max}$} & $\Delta_{\mathcal{P}}$ & $F_{\mathrm{HS}}$ & $D_{\mathrm{HS}}$ \\
    \hline
    unconstrained            &  {6} & 0.000 & 1.000 & 0.003 \\
    $1|5$                    &  {5} & 0.464 & 0.245 & 1.229 \\
    $2|4$                    &  {4} & 0.569 & 0.000 & 1.414 \\
    $3|3$                    &  {3} & 0.598 & 0.500 & 1.000 \\
    $2|2|2$                  &  {2} & 1.037 & 0.038 & 1.387 \\
    $1|1|1|1|1|1$            &  {1} & 2.006 & 0.118 & 1.328 \\
    \hline\hline
  \end{tabular}
  \caption{
Training results for the six-qubit GHZ target,
    Eq.~\eqref{eq:psi_A}.  For each ansatz we report the maximal block
    size $d_{\max}$, the minimal NLL gap $\Delta_{\mathcal{P}}$ with
    respect to the global (unrestricted) model, the Hilbert--Schmidt
    fidelity $F_{\mathrm{HS}}$, and the Hilbert-Schmidt distance, $D_{\mathrm{HS}}$, between the
    reconstructed and exact states.  Rows are ordered by increasing
    $\Delta_{\mathcal{P}}$.
  }
  \label{tab:A_pure}
\end{table}

\begin{table}[t]
  \centering
  \begin{tabular}{|c|c|c|c|c|}
    \hline\hline
    Partition $\mathcal{P}$ & $d_{\max}$ & $\Delta_{\mathcal{P}}$ & $F_{\mathrm{HS}}$ & $D_{\rm HS}$ \\
    \hline
    unconstrained                    & 6 & 0.000 & 1.000 & $< 10^{-2}$ \\
    $2|4$                     & 4 & 0.000 & 1.000 & $< 10^{-2}$ \\
    $2|2|2$                   & 2 & 0.000 & 1.000 & $< 10^{-2}$ \\
    $3|3$                     & 3 & 0.636 & 0.173 & 1.286 \\
    $1|5$                     & 5 & 0.649 & 0.499 & 1.001 \\
    $1|1|1|1|1|1$             & 1 & 1.920 & 0.024 & 1.397 \\
    \hline\hline
  \end{tabular}
  \caption{
    Training results for the product of three Bell pairs,
    Eq.~\eqref{eq:psi_B}.  Columns as in
    Table~\ref{tab:A_pure}.  Rows are ordered by increasing
    $\Delta_{\mathcal{P}}$.
  }
  \label{tab:B_pure}
\end{table}
The second example is a six-qubit state consisting of three independent
Bell pairs,
\begin{equation}
  \label{eq:psi_B}
  \ket{\psi_{\mathrm{B}}}
  \equiv
  \ket{\Phi^+}^{\otimes 3},
  \qquad
  \ket{\Phi^+}
  = \frac{1}{\sqrt{2}}\bigl(\ket{0_z 0_z} + \ket{1_z 1_z}\bigr).
\end{equation}
This state has entanglement depth $d_e(\ket{\psi_{\mathrm{B}}}) = 2$ and
exhibits only short-range two-qubit correlations.
We again train the global NQS together with the same family of SNQS
partitions as in the GHZ benchmark.

The likelihood hierarchy now changes qualitatively.
Taking the global ansatz as reference
($\Delta_{\mathrm{global}} = 0$), we obtain
\[
  \Delta_{2|2|2} \approx 0,\qquad
  \Delta_{2|4} \approx 0,
\]
within numerical precision, while partitions that break the Bell-pair
structure incur substantial positive gaps,
\[
  \Delta_{3|3} \approx 0.64,\qquad
  \Delta_{1|5} \approx 0.65,\qquad
  \Delta_{1|1|1|1|1|1} \approx 1.92.
\]
The ansatz with partition $2|2|2$ (three two-qubit blocks) matches the
performance of the global model, as visible in
Fig.~\ref{fig:fig2}(b), reflecting the fact that the target
is exactly $2$-producible with respect to the natural Bell-pair
partition.  
The partition $2|4$ also achieves essentially zero gap: the four-qubit
block is flexible enough to internally represent two Bell pairs, so the
target again lies within the corresponding model family.

From the perspective of entanglement depth, these results show that all
 {$l=1$}--producible (fully separable) SNQS models are strongly disfavoured
by the data, whereas there exist  {$l=2$}--producible architectures
($d_{\max}=2$) whose NLL is indistinguishable from the global model.
Consequently, the likelihood--gap criterion certifies
\[
  d_e(\ket{\psi_{\mathrm{B}}}) > 1,
\]
and the smallest  {$l$} for which the data admit a compatible description
is  {$l=2$}, consistent with the exact entanglement depth.
As in the GHZ case, the architectures with vanishing NLL gap yield
reconstructed density matrices with $F_{\mathrm{HS}}\approx 1$, while
models with large gaps exhibit visibly reduced HS fidelities and HS
distances close to unity (Table~\ref{tab:B_pure}).

\subsubsection{Ten-qubit product of entangled clusters}
\label{subsec:ten-qubit-product}

\begin{table}[t]
  \centering
  \begin{tabular}{|c|c|c|c|c|}
    \hline\hline
    Partition $\mathcal{P}$ & $d_{\max}$ & $\Delta_{\mathcal{P}}$ & $F_{\mathrm{HS}}$ & $D_{\rm HS}$ \\
    \hline
    unconstrained                    & 10 & 0.000 & 1.000 & $2.9 \times 10^{-2}$ \\
    $6|4$                     &  6 & 0.000 & 1.000 & $1.5 \times 10^{-2}$ \\
    $3|7$                     &  7 & 0.000 & 1.000 & $1.6 \times 10^{-2}$ \\
    $3|3|4$                   &  4 & 0.000 & 1.000 & $1.7 \times 10^{-2}$ \\
    $4|6$                     &  6 & 0.443 & 0.518 & 0.981 \\
    $4|2|4$                   &  4 & 0.444 & 0.601 & 0.893 \\
    $5|1|4$                   &  5 & 0.479 & 0.575 & 0.921 \\
    $3|4|3$                   &  4 & 0.530 & 0.280 & 1.200 \\
    $7|3$                     &  7 & 0.531 & 0.043 & 1.383 \\
    $2|4|4$                   &  4 & 0.559 & 0.264 & 1.213 \\
    $1|5|4$                   &  5 & 0.592 & 0.246 & 1.228 \\
    $5|5$                     &  5 & 0.647 & 0.008 & 1.409 \\
    $4|1|5$                   &  5 & 0.654 & 0.010 & 1.407 \\
    $4|3|3$                   &  4 & 0.904 & 0.114 & 1.331 \\
    $4|5|1$                   &  5 & 0.963 & 0.045 & 1.382 \\
    $1|4|5$                   &  5 & 1.006 & 0.239 & 1.234 \\
    $4|4|2$                   &  4 & 1.021 & 0.252 & 1.224 \\
    $1|8|1$                   &  8 & 1.074 & 0.140 & 1.312 \\
    $5|4|1$                   &  5 & 1.140 & 0.009 & 1.408 \\
    $2|2|2|2|2$               &  2 & 1.458 & 0.019 & 1.401 \\
    $1|1|1|1|1|1|1|1|1|1$     &  1 & 21.736 & 0.031 & 1.000 \\
    \hline\hline
  \end{tabular}
  \caption{
 Training results for the ten-qubit state,
    Eq.~\eqref{eq:psi_C}.  Columns as in
    Table~\ref{tab:A_pure}.  Rows are ordered by increasing
    $\Delta_{\mathcal{P}}$.
  }
\label{tab:C_pure}
\end{table}

The final benchmark is a ten-qubit pure state composed of three
entangled clusters,
\begin{equation}\label{eq:psi_C}
  \ket{\psi_{\mathrm{C}}}
  =
  \ket{\psi_{\mathrm{GHZ}}^{\mathrm{XYZ}}}
  \otimes
  \ket{\mathrm{D}^{(3)}_{1}}
  \otimes
  \ket{\psi_{\mathrm{GHZ}}^{\mathrm{XZXY}}} ,
\end{equation}
where the superscripts indicate local single-qubit basis rotations, and
$\ket{\mathrm{D}^{(N)}_{n}}$ denotes a Dicke state with $n$ excitations
among $N$ qubits,
\begin{equation}
  \ket{\mathrm{D}^{(N)}_{n}}
  =
  \binom{N}{n}^{-1/2}
  \sum_{\pi} P_{\pi}
  \bigl(
    \ket{0}^{\otimes (N-n)} \otimes \ket{1}^{\otimes n}
  \bigr),
\end{equation}
with the sum taken over all distinct permutations $P_{\pi}$.
The target thus consists of a three-qubit GHZ state in $(X,Y,Z)$ bases,
a three-qubit Dicke state with one excitation, and a four-qubit GHZ
state in $(X,Z,X,Y)$ bases.
Its entanglement depth is set by the largest entangled cluster,
\begin{equation}
  d_e(\ket{\psi_{\mathrm{C}}}) = 4,
\end{equation}
while the global structure is a tensor product across the three
clusters.

To probe how this heterogeneous structure is reflected in likelihood
gaps, we train the global NQS together with a broad family of SNQS
partitions on random-Pauli data for $N=10$.  
The partitions include two-block and three-block architectures such as
$4|6$, $6|4$, $7|3$, $3|7$, $5|5$, $3|3|4$, $4|3|3$, $3|4|3$,
$4|2|4$, $2|4|4$, $4|4|2$, $5|1|4$, $4|1|5$, $1|5|4$, $1|8|1$, as well
as strongly factorized models $2|2|2|2|2$ and the fully product ansatz
$1|1|1|1|1|1|1|1|1|1$.  
Representative training curves are shown in
Fig.~\ref{fig:fig2}(c), and the optimized losses and HS
diagnostics are listed in
Table~\ref{tab:C_pure}.

The global, fully entangling NQS sets the reference value
$\Lstarfull$ and achieves $F_{\mathrm{HS}}\approx 1$ with a very small
HS distance.
Strikingly, several structured ansatzes with maximal block size
$d_{\max} \ge 4$ attain essentially indistinguishable NLL and
$F_{\mathrm{HS}}\approx 1$:
the partitions $6|4$, $3|7$, and, most importantly, the physically
natural $3|3|4$ all have $\Delta_{\mathcal{P}}\simeq 0$.
Although these architectures group the ten qubits in different ways,
none of them enforce an entanglement depth smaller than $4$, and all
are expressive enough to reproduce the data at the level of finite-shot
statistics.
From the point of view of our protocol, they are therefore all
compatible with the observed measurement outcomes.

By contrast, partitions that fragment the physical clusters or impose
an incompatible factorization structure exhibit sizeable positive
likelihood gaps.
Examples include
\[
  \Delta_{4|6} \approx 0.44,\ 
  \Delta_{4|2|4} \approx 0.44, 
  \Delta_{5|1|4} \approx 0.48,
  \Delta_{3|4|3} \approx 0.53,
\]
all accompanied by a noticeable drop in $F_{\mathrm{HS}}$ and an
increase in HS distance.
Here, the SNQS partition forces the model either to cut through at
least one of the entangled clusters or to entangle qubits that are
unentangled in the true tensor-product structure; in both cases the
resulting model family cannot simultaneously accommodate all empirical
correlations, and the NLL gap becomes a visible ``entanglement
penalty''.

This trend is amplified for strongly over-fragmented ansatzes.
Architectures with small $d_{\max}$, such as $2|2|2|2|2$ and the fully
product partition $1|1|1|1|1|1|1|1|1|1$, show gaps of order unity and
well beyond (e.g.\ $\Delta_{2|2|2|2|2} \approx 1.46$ and
$\Delta_{1|1|\dots|1} \approx 21.7$), together with very low
Hilbert--Schmidt fidelities and HS distances close to their maximal
values in our normalization.
These models are manifestly unable to reproduce the measurement
statistics of a state with genuine four-qubit entanglement.

Two robust conclusions follow from this benchmark.
First, among all tested SNQS architectures, every partition with
$d_{\max} \le 3$ (namely $2|2|2|2|2$ and $1|1|1|1|1|1|1|1|1|1$)
exhibits a substantial positive gap $\Delta_{\mathcal{P}}$, so the data
are incompatible with any  {$l$}-producible model with  {$l \le 3$} within
our ansatz hierarchy.
Second, several architectures with $d_{\max} = 4$ or larger (in
particular $3|3|4$) reproduce the likelihood of the global model up to
finite-shot fluctuations, showing that an entanglement depth of $4$ is
sufficient to describe the system.
Combining these observations, the likelihood--gap criterion certifies
\[
  d_e(\ket{\psi_{\mathrm{C}}}) > 3,
\]
in agreement with the known value $d_e = 4$, and it does so using only
NLL comparisons between neural ansatzes trained on measurement data,
without full tomographic reconstruction of the ten-qubit state.

\vspace{0.5em}
Across all three pure-state benchmarks a coherent picture emerges:
whenever the SNQS partition $\mathcal{P}$ is too restrictive to capture
the true entanglement structure (i.e.\ it enforces a maximal block size
$d_{\max}(\mathcal{P})$ smaller than the actual entanglement depth),
training produces a robust positive gap $\Delta_{\mathcal{P}} \gg 0$;
conversely, as soon as the hierarchy includes an ansatz with
$d_{\max}(\mathcal{P})$ at least as large as the true entanglement
depth, the corresponding likelihood becomes indistinguishable from that
of the global model.
This is precisely the behaviour required for likelihood-based
entanglement-depth certification without full tomography.

}

\subsection{Mixed states}
\label{sec:results-mixed}

We now move beyond ideal pure states and address the certification of
entanglement depth in more realistic scenarios where decoherence maps
the target \(\ket{\psi_{\mathrm{tgt}}}\) onto a mixed density matrix
\(\rho_{\mathrm{tgt}}\).
In this regime a useful witness must distinguish genuine quantum
entanglement from classical statistical correlations generated by
incoherent noise.
We demonstrate that the likelihood--gap criterion continues to work in
this setting by reusing the three target states from the previous section,
now subject to a noise channel.

To mimic energy relaxation in, e.g., superconducting qubit platforms, we
model the environment by independent amplitude-damping channels acting
on each qubit.
For a single qubit with density matrix \(\rho\) this channel is defined
by the Kraus operators
\begin{equation}
  K_0
  =
  \begin{pmatrix}
    1 & 0 \\
    0 & \sqrt{1-p}
  \end{pmatrix},
  \qquad
  K_1
  =
  \begin{pmatrix}
    0 & \sqrt{p} \\
    0 & 0
  \end{pmatrix},
\end{equation}
through the completely positive trace-preserving map
\begin{equation}
  \mathcal{E}_{\mathrm{AD}}(\rho)
  =
  K_0 \rho K_0^\dagger
  + K_1 \rho K_1^\dagger.
\end{equation}
Physically, \(K_0\) leaves \(\ket{0}\) unchanged and damps the
excited state \(\ket{1}\) towards \(\ket{0}\), while \(K_1\) describes a
stochastic decay \(\ket{1}\to\ket{0}\) with probability \(p\).
For an \(N\)-qubit state we apply \(\mathcal{E}_{\mathrm{AD}}\) to each
qubit independently,
\begin{equation}
  \rho_{\mathrm{tgt}}
  =
  \bigl(
    \mathcal{E}_{\mathrm{AD}}^{\otimes N}
  \bigr)
  \bigl[
    \ket{\psi_{\mathrm{tgt}}}
    \!\bra{\psi_{\mathrm{tgt}}}
  \bigr],
\end{equation}
which drives the system towards the separable ground state
\(\ket{0}^{\otimes N}\) and reduces the purity
\(\Tr(\rho_{\mathrm{tgt}}^2)\).

To accommodate mixed targets we generalize the variational ansatz from a
single pure neural quantum state to a small \textit{ensemble} of pure
NQS components.
For the global (unstructured) model we write
\begin{equation}
  \rho_{\mathrm{glob}}
  =
  \sum_{r=1}^{R}
  w_r\,
  \ket{\psi_r}\!\bra{\psi_r},\qquad
  w_r \ge 0,\quad
  \sum_r w_r = 1,
\end{equation}
where each \(\ket{\psi_r}\) is a pure NQS with its own parameters and
the \(w_r\) are learnable mixing weights.
In all experiments below we use a small rank \(R=4\).

For a structured mixed ansatz compatible with a partition
\(\mathcal{P} = \{C_1,\dots,C_B\}\) we impose the SNQS product structure
inside \textit{each} component of the ensemble.
The corresponding density matrix is
\begin{equation}
  \rho_{\mathcal{P}}
  =
  \sum_{r=1}^{R}
    w_r \;
    \bigotimes_{C\in\mathcal{P}}
      \ket{\psi^{(r)}_C}\!\bra{\psi^{(r)}_C},
  \label{eq:ensemble-SNQS}
\end{equation}
where \(\ket{\psi^{(r)}_C}\) is a pure NQS defined only on block
\(C\subset\{1,\dots,N\}\).
For each component \(r\), the global state factorizes exactly across the blocks in \(\mathcal{P}\), i.e.
there is \textit{no} entanglement between different blocks in a  single component. The ensemble weights \(w_r\) introduce a hidden classical
 randomness: before each (virtual) experimental shot the model ``chooses'' one of the product patterns \(r\) and prepares that configuration.

For a given partition \(\mathcal{P}\) the ensemble thus spans a rich
family of mixed states that are separable with respect to
\(\mathcal{P}\), in the sense that they can be written as a classical
mixture of block-product pure states.
Such states can exhibit complicated  classical correlations
between blocks, but they cannot reproduce
quantum correlations across the cut.
Within this framework the likelihood gap
\(\Delta_{\mathcal{P}}\) retains a
simple operational meaning:
(i) if \(\Delta_{\mathcal{P}}\approx 0\), then the measurement data are compatible with such a classical lottery over product  configurations; no entanglement across \(\mathcal{P}\) is needed to fit the data within our ansatz family, (ii) if \(\Delta_{\mathcal{P}}\gg 0\), then \textit{even allowing for classical mixtures of many block-product states} the model  cannot reproduce the observed statistics.
The remaining correlations across the cut must therefore be  genuinely quantum in origin.

We start with six-qubit GHZ state of
Eq.~\eqref{eq:psi_A} under local amplitude damping.
The corresponding results are summarized in
Table~\ref{tab:A_mixed}.
\begin{table}[t!]
  \centering
  \begin{tabular}{|c|c|c|c|c|}
    \hline\hline
    Partition $\mathcal{P}$ & $d_{\max}$ & $\Delta_{\mathcal{P}}$ & $F_{\mathrm{HS}}$ & $D_{\mathrm{HS}}$ \\
    \hline
    unconstrained                    & 6 & 0.000 & 0.997 & 0.067 \\
    $1|5$                     & 5 & 0.026 & 0.723 & 0.600 \\
    $2|4$                     & 4 & 0.029 & 0.711 & 0.611 \\
    $3|3$                     & 3 & 0.029 & 0.709 & 0.613 \\
    $2|2|2$                   & 2 & 0.030 & 0.704 & 0.618 \\
    $1|1|1|1|1|1$             & 1 & 0.030 & 0.704 & 0.619 \\
    \hline\hline
  \end{tabular}
  \caption{
    Training results for the mixed six-qubit GHZ state, Eq.\eqref{eq:psi_A}, under amplitude damping channel
    ($p=0.05$).
    Despite the loss of purity, all structured ensembles with
    $d_{\max} < 6$ exhibit systematic positive likelihood gaps, showing
    that the data cannot be fully explained by mixtures of
    product-structured components with bounded entanglement depth.
  }
  \label{tab:A_mixed}
\end{table}
Compared to the pure-state case the absolute sizes of the gaps are
reduced by roughly an order of magnitude: all structured ensemble
ansatzes lie within \(\Delta_{\mathcal P}\sim 3\times 10^{-2}\) of the
global model.
This is expected, as amplitude damping drives the state partially
towards the separable ground state and hence weakens the
multi-qubit coherences.
However, a clear pattern remains: the global ensemble achieves the
lowest NLL and high Hilbert--Schmidt fidelity
(\(F_{\mathrm{HS}}\approx 0.997\)), while \textit{every} partition with
\(d_{\max}<6\) incurs a consistent positive penalty.

Within the ensemble SNQS picture this means that, even after allowing
for classical mixtures of block-product NQS, no model with maximal block
size \(d_{\max}\le 5\) can fully account for the noisy GHZ data.
The mixed target is therefore incompatible with any  {$l$}-producible
ensemble SNQS with  {$l\le 5$} in our hierarchy, and we again certify
\[
  d_e(\rho_{\mathrm{GHZ}_6}^{\mathrm{(mixed)}}) > 5,
\]
i.e.\ an entanglement depth of at least six qubits survives on top of
the amplitude-damping noise.

The second mixed benchmark starts from three independent Bell pairs,
Eq.~\eqref{eq:psi_B}, and subjects them to the same local
amplitude-damping channel.
This example is particularly informative because it contains both
\textit{short-range} entanglement within each pair and potential
\textit{classical} correlations between pairs induced by noise.
The results are listed in Table~\ref{tab:B_mixed}.

\begin{table}[t]
  \centering
  \begin{tabular}{|c|c|c|c|c|}
    \hline\hline
    Partition $\mathcal{P}$ & $d_{\max}$ & $\Delta_{\mathcal{P}}$ & $F_{\mathrm{HS}}$ & $D_{\mathrm{HS}}$\\
    \hline
    unconstrained                    & 6 & 0.000 & 0.998 & 0.061 \\
    $2|4$                     & 4 & 0.000 & 0.998 & 0.060 \\
    $2|2|2$                   & 2 & 0.001 & 0.998 & 0.059 \\
    $3|3$                     & 3 & 0.099 & 0.806 & 0.523 \\
    $1|5$                     & 5 & 0.106 & 0.812 & 0.518 \\
    $1|1|1|1|1|1$             & 1 & 0.466 & 0.265 & 0.876 \\
    \hline\hline
  \end{tabular}
  \caption{
     Training results for mixed state of three Bell pairs,
    Eq.~\eqref{eq:psi_B} under amplitude damping channel ($p=0.05$).
  }
  \label{tab:B_mixed}
\end{table}

The likelihood hierarchy follows the physical structure of the state.
The partitions $2|2|2$ and $2|4$, which respect or coarse-grain the
Bell-pair grouping, achieve NLL values indistinguishable from the global
ensemble and HS fidelities \(F_{\mathrm{HS}}\approx 0.998\).
This shows that any correlations \textit{between} Bell pairs created by
the noise can be modeled as classical correlations between different
product patterns of entangled two-qubit blocks.

In stark contrast, partitions that cut through at least one Bell pair
($3|3$ and $1|5$) incur clear positive gaps
($\Delta_{\mathcal P}\approx 0.1$) and substantially reduced
Hilbert--Schmidt fidelities.
The fully product ansatz $1|1|\dots|1$ performs worst, as expected, with
\(\Delta\approx 0.47\) and  low fidelity.
Since the $3|3$ ansatz already has access to a rank-4 ensemble of
product states across the cut, its failure indicates that the correlations within each Bell pair
cannot be decomposed into a mixture of product states, and must
therefore be genuinely quantum.

Finally, we consider ten-qubit product state composed of GHZ and
Dicke clusters discussed in Eq.~\eqref{eq:psi_C}, subjected to the same noise channel.
Table~\ref{tab:C_mixed} reports the likelihood gaps and
Hilbert--Schmidt diagnostics for the same family of partitions as in the
pure-state case.

\begin{table}[t!]
  \centering
  \begin{tabular}{|c|c|c|c|c|}
    \hline\hline
    Partition $\mathcal{P}$ & $d_{\max}$ & $\Delta_{\mathcal{P}}$ & $F_{\mathrm{HS}}$ & $D_{\mathrm{HS}}$ \\
    \hline
   unconstrained                    & 10 & 0.000 & 0.987 & 0.137 \\
    $6|4$                     & 6  & 0.001 & 0.989 & 0.125 \\
    $3|7$                     & 7  & 0.044 & 0.836 & 0.440 \\
    $3|3|4$                   & 4  & 0.066 & 0.765 & 0.516 \\
    $7|3$                     & 7  & 0.066 & 0.663 & 0.599 \\
    $2|4|4$                   & 4  & 0.075 & 0.682 & 0.585 \\
    $1|5|4$                   & 5  & 0.082 & 0.647 & 0.612 \\
    $3|4|3$                   & 4  & 0.105 & 0.581 & 0.652 \\
    $1|8|1$                   & 8  & 0.117 & 0.509 & 0.695 \\
    $4|6$                     & 6  & 0.189 & 0.572 & 0.660 \\
    $5|5$                     & 5  & 0.209 & 0.531 & 0.687 \\
    $4|4|2$                   & 4  & 0.242 & 0.452 & 0.729 \\
    $5|1|4$                   & 5  & 0.247 & 0.469 & 0.720 \\
    $1|4|5$                   & 5  & 0.317 & 0.383 & 0.768 \\
    $4|2|4$                   & 4  & 0.335 & 0.384 & 0.764 \\
    $4|5|1$                   & 5  & 0.343 & 0.352 & 0.781 \\
    $5|4|1$                   & 5  & 0.345 & 0.358 & 0.778 \\
    $4|1|5$                   & 5  & 0.355 & 0.343 & 0.789 \\
    $4|3|3$                   & 4  & 0.366 & 0.320 & 0.797 \\
    $2|2|2|2|2$               & 2  & 0.427 & 0.281 & 0.816 \\
    $1|1|\dots|1$             & 1  & 21.55 & 0.039 & 0.799 \\
    \hline\hline
  \end{tabular}
  \caption{
   Training results for the ten-qubit state,
    Eq.~\eqref{eq:psi_C} under amplitude damping channel ($p=0.05$).  Columns as in
    Table~\ref{tab:A_pure}.
  }
  \label{tab:C_mixed}
\end{table}

  {Note that the partition $7|3$ shows a gap ($\Delta_{7|3}\approx 0.066$) comparable to $3|3|4$, despite cutting through the four-qubit GHZ cluster. This coincidence is a noise effect: under amplitude damping ($p=0.05$), the four-qubit GHZ coherences are partially suppressed, so the NLL penalty from fragmenting this block is comparable to the penalty incurred by the ensemble's limited rank ($R=4$) in the $3|3|4$ fit. In contrast, in the pure-state case (Table~\ref{tab:C_pure}), $7|3$ exhibits a substantially larger gap ($\Delta_{7|3}\approx 0.53$) while $3|3|4$ achieves $\Delta\approx 0$, clearly distinguishing the two. The protocol's certification of $d_e>3$ remains correct in both cases.}

The unconstrained NQS and the $6|4$ partition achieve nearly identical
likelihoods and high Hilbert--Schmidt fidelities, indicating that
grouping the first two clusters into a single effective block is
compatible with the data.
The natural tripartition $3|3|4$ (\(\Delta_{3|3|4}\approx 0.064\)), and bipartition $3|7$ (\(\Delta_{3|7}\approx 0.044\)) which follow the true cluster
boundaries, show small gaps, attributable to the limited
ensemble rank when modeling three independent mixed clusters.
 {Scaling the ensemble rank $R$ is necessary for highly mixed states and represents a hyperparameter trade-off in the current implementation.}

More importantly, partitions that attempt to fragment the four-qubit GHZ
cluster, such as $4|3|3$ and $5|5$, incur substantially larger gaps
(\(\Delta\approx 0.35\)--\(0.4\)) and markedly reduced HS fidelities.
Strongly over-fragmented partitions like $2|2|2|2|2$ and the fully
product $1|1|\dots|1$ perform worst, as expected.

From the point of view of entanglement depth, the data are incompatible
with any ensemble SNQS whose maximal block size satisfies
\(d_{\max}\le 3\), while several partitions with \(d_{\max}=4\) or
larger match the performance of the global ensemble within statistical
uncertainty.
Our likelihood--gap criterion therefore certifies
$d_e(\rho_{\mathrm{C}}^{\mathrm{(mixed)}}) > 3$,
consistent with the four-qubit GHZ cluster present in the underlying
pure state.

At the same time, the ten-qubit example illustrates an important
feature of the method: when multiple partitions with \(d_{\max}\ge 4\)
achieve very similar likelihoods (e.g.\ $6|4$, $3|7$, $3|3|4$), the
NLL landscape does not single out a unique block structure.
In this regime our protocol correctly refrains from over-interpreting
the data: it reliably detects that some entangled cluster of size at
least four is required, but it does not artificially pin down a
particular partition when the evidence is ambiguous.

For mixed states we test separability with respect to fixed partitions 
${\cal P}$
using low-rank mixtures of
${\cal P}$-product components; allowing mixtures across different
${\cal P}$ would further enlarge the constrained hypothesis class and can be incorporated in future work.
Our conclusions are variational with respect to the chosen SNQS model class and the tested set of partitions; large likelihood gaps rule out those hypotheses within this modeling family.

\section{Interpretability of Unrestricted Neural Quantum States}
\label{sec:interpretability}

While the likelihood-gap protocol treats the unrestricted ansatz $\mathcal{A}_{full}$ primarily as a flexible reference for model comparison, it is natural to ask what physical information is encoded in its trained parameters. In physics applications, accurate likelihood fits are most useful when they can be accompanied by diagnostics that provide sanity checks on learned structure and yield compact, human-interpretable summaries that can guide subsequent modeling choices. A number of approaches along these lines have been developed, including tools for interpreting neural decision functions and assessing reliability~\cite{Dawid2020,Dawid2021}, interpretable-by-construction architectures and order-parameter extractors that expose correlation structure~\cite{Wetzel2017,Miles2021,Valenti2022,Cybinski2024,Doeschl2025}, and sparsification or weight-space analyses that connect salient parameters and generalization behavior to underlying physical organization~\cite{Golubeva2022,Barton2025,Hernandes2025,Moss2025}. In the present setting, we show that simple post-processing of the unrestricted NQS weights yields qualitative signatures of subsystem and entanglement structure directly at the level of learned parameters.

We begin with the characterization of pure target states with the help of two standard tools. First, we  consider a two-body correlator
\begin{equation}
  c_{ij}^{\alpha\beta}
  =
  \langle \sigma_i^\alpha \sigma_j^\beta \rangle
  -
  \langle \sigma_i^\alpha \rangle
  \langle \sigma_j^\beta \rangle,
  \qquad
  i\neq j,
  \label{eq:connected-correlators}
\end{equation}
$\alpha,\beta\in\{X,Y,Z\}$, and define its basis aggregation
\begin{equation}
  C_{ij}
  =
  \Biggl(
    \sum_{\alpha,\beta\in\{X,Y,Z\}}
      \bigl(c_{ij}^{\alpha\beta}\bigr)^2
  \Biggr)^{1/2}.
  \label{eq:Cij-def}
\end{equation}
The \(C_{ij}\) correlator can be extracted either from the quantum state, directly from raw measurement bitstrings, or from trained NQS. 
In randomized-basis datasets, a direct estimator of each component
$c_{ij}^{\alpha\beta}$ uses only those experimental shots in which qubits
$i$ and $j$ were measured in the corresponding local bases
$(\alpha,\beta)$.
For finite measurement budgets this results in an estimator whose
variance is both basis-dependent and {uneven across pairs}
$(i,j)$ and settings $(\alpha,\beta)$, because different basis patterns
occur with different frequencies in the dataset.
A trained generative model provides an alternative, {model-based}
estimate of the same correlators: once the NQS is fitted by maximum
likelihood to the full dataset, expectation values can be evaluated under
the learned state using all measurement settings simultaneously, which can reduce estimator variance at the cost of model bias, analogous to standard regularized or model-based estimators.

\begin{figure}[t]
  \centering
  \includegraphics[width=0.99\linewidth]{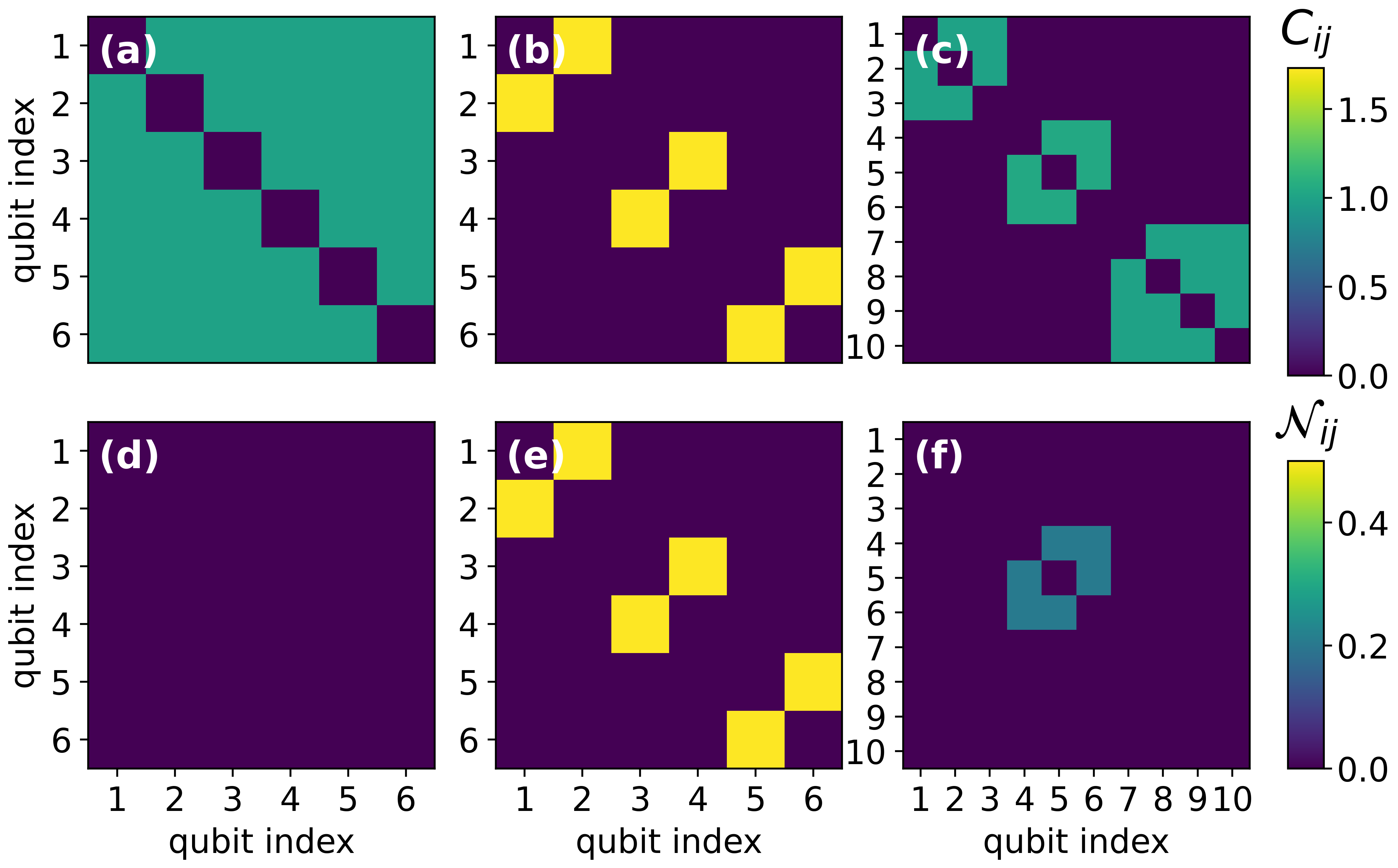}
  \caption{%
   Aggregated two-body correlator \(C_{ij}\), Eq.~\eqref{eq:Cij-def} (top row), and negativity, Eq.~\eqref{eq:negativity_eigs} (bottom row),
    for the three benchmark states: six-qubit GHZ state
    \(\ket{\psi_{\mathrm{A}}}\), Eq.~\eqref{eq:psi_A}, product of three Bell pairs 
    \(\ket{\psi_{\mathrm{B}}}\), Eq.~\eqref{eq:psi_B}, and ten-qubit product of three entangled clusters \(\ket{\psi_{\mathrm{C}}}\), Eq.~\eqref{eq:psi_C}.
  }
  \label{fig:Cij-maps}
\end{figure}
We additionally quantify two-qubit entanglement via the {negativity}. For each qubit pair $(i,j)$ we consider the reduced density matrix $\rho_{ij}=\mathrm{Tr}_{\overline{ij}}(\rho)$ and take its partial transpose with respect to one subsystem (here $j$), denoted $\rho_{ij}^{T_j}$. The negativity is then defined as
\begin{equation}
  \mathcal{N}_{ij}
  =
  \sum_{\lambda_k<0} |\lambda_k|,
  \label{eq:negativity_eigs}
\end{equation}
where $\{\lambda_k\}$ are the eigenvalues of $\rho_{ij}^{T_j}$.
 We consider negativity as a reference entanglement diagnostic. In the
synthetic benchmarks  here it is computed from the exact target state; in an experimental setting it could
only be accessed through a reconstructed model state via the trained
NQS.

Figure~\ref{fig:Cij-maps} shows \(C_{ij}\) (top row) and ${\cal N}_{ij}$ (bottom row) for considered pure states $\ket{{\psi}_{\rm A,B,C}}$ (columns from left to right, respectively).
For six-qubit GHZ state, $\ket{\psi_{\rm A}}$ (panel (a)), $C_{ij}$ displays broadly distributed two--body
correlations, reflecting its global symmetry, while negativity (panel (d)) vanishes, as the GHZ state is genuinely entangled state, and after tracing out any qubit, the reduced density matrix has no entanglement. 
For the product of
three Bell pairs \(\ket{\psi_{\mathrm{B}}}\) (panel (b)), \(C_{ij}\) consists of three
isolated strong links, sharply revealing the \(2|2|2\) cluster structure. The negativity (panel (e)) displays similar structure, showing strong entanglement between qubits in each of the Bell pairs.
For the heterogeneous ten--qubit state, \(\ket{\psi_{\mathrm{C}}}\),
\(C_{ij}\) reveals a block pattern consistent with the intended
\(3|3|4\) factorization (panel (c)). The negativity (panel (f)) shows strong entanglement within central 3-qubit block of the $\ket{\textrm{D}^{(3)}_1}$ Dicke state, and vanishes for blocks of GHZ states.

\subsection{Weight--space diagnostics of the unrestricted NQS}
\label{subsec:interpret-weights}

We now turn to structure encoded directly in the trained RBM weights.
For a spin configuration \(\mathbf{s}\in\{\pm 1\}^N\) we write the NQS
wavefunction as
\begin{equation}
  \Psi_\theta(\mathbf{s})
  =
  \exp\!\bigl[
    A_\theta(\mathbf{s}) + i \Phi_\theta(\mathbf{s})
  \bigr],
\end{equation}
where \(A_\theta\) and \(\Phi_\theta\) are represented by RBMs of the form
\begin{equation}
  X_\theta(\mathbf{s})
  =
  \sum_i x_i s_i
  + \sum_{h}
      \log\!\left[
        2\cosh\!\Bigl(
          y_h + \sum_i W^{(X)}_{ih} s_i
        \Bigr)
      \right],
  \label{eq:rbm-X-generic}
\end{equation}
where  $X\in\{A,\Phi\}$. In what follows we focus on the visible--hidden weight matrices
\(W^{(A)}\) and \(W^{(\Phi)}\), and construct from them two
\(N\times N\) summaries that can be compared to the correlator \(C_{ij}\).

\begin{figure}[t]
  \centering
  \includegraphics[width=0.99\linewidth]{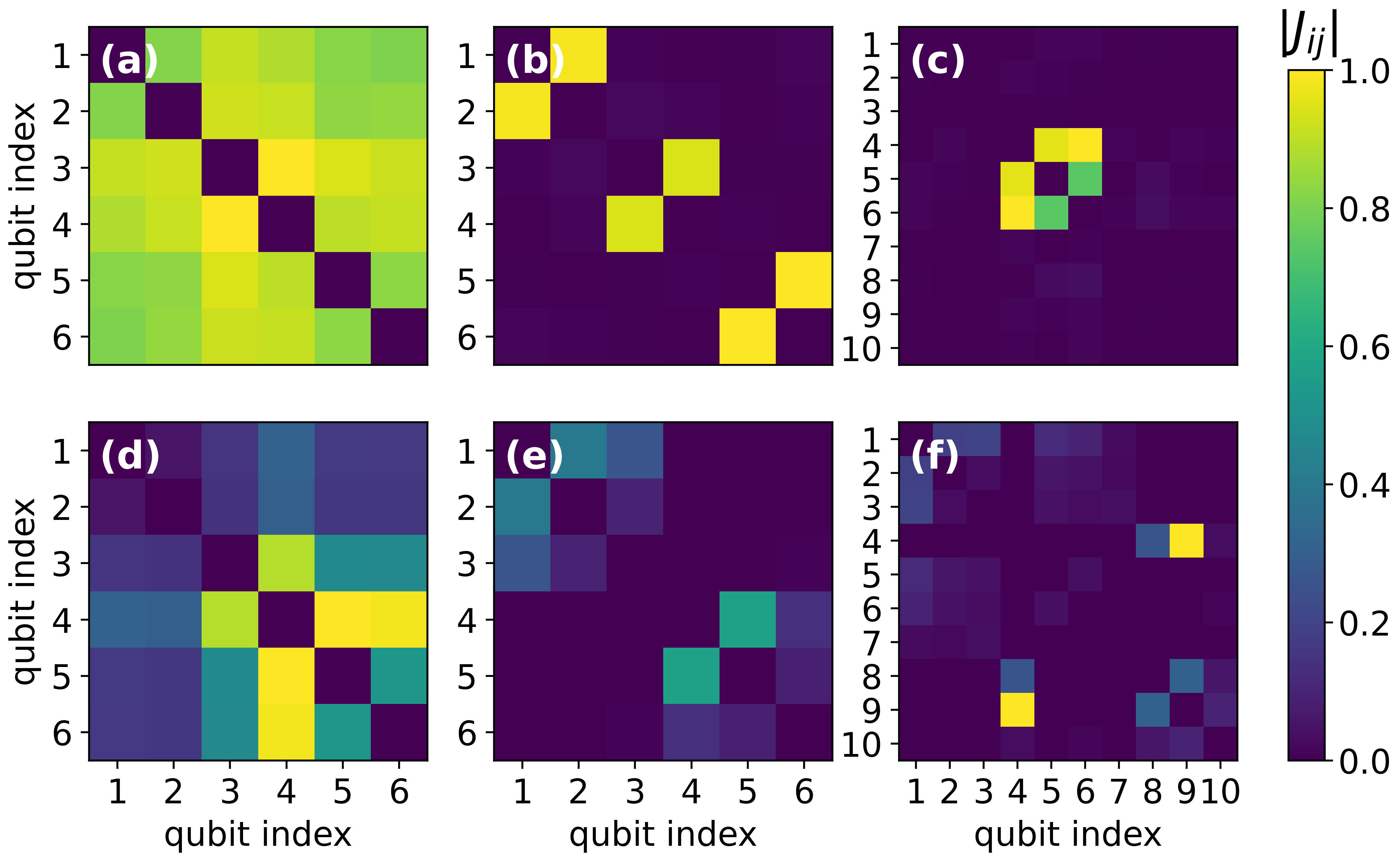}
  \caption{%
    Effective visible nodes (qubits) couplings extracted from the trained weight $W$ of an unrestricted
    NQS.  Each panel shows the absolute value of the normalized coupling
    matrix \(|J^{(X)}_{ij}|\), with \(X=A\) (amplitude, top row) or
    \(X=\Phi\) (phase, bottom row), computed from the Gram matrices
    \(J^{(X)} = W^{(X)} {W^{(X)}}^{\!\!T}\) of the trained unconstrained NQS.
    Columns correspond to the three pure targets : (a,d) six-qubit GHZ
    \(\ket{\psi_{\mathrm{A}}}\), Eq.\eqref{eq:psi_A}, (b,e) product of three Bell pairs
    \(\ket{\psi_{\mathrm{B}}}\), Eq.\eqref{eq:psi_B}, and (c,f) ten-qubit product state with
    GHZ and Dicke clusters \(\ket{\psi_{\mathrm{C}}}\), Eq.\eqref{eq:psi_C}.  
    For \(\ket{\psi_{\mathrm{A}}}\) the couplings
    are comparatively homogeneous, reflecting the genuinely global
    character of the GHZ entanglement.
    For
    \(\ket{\psi_{\mathrm{B}}}\) the amplitude couplings display three
    localized \(2\times 2\) blocks, directly exposing the underlying
    \(2|2|2\) structure.  For
    \(\ket{\psi_{\mathrm{C}}}\)
    the dominant amplitude couplings concentrate on the three-qubit Dicke cluster, consistent with the fact that pairwise entanglement is present only in that block (Fig.~\ref{fig:Cij-maps}f).
    The rotated bases GHZ states are not simply visible, reflecting complicated structure in the computational basis, natural for $J^{(X)}_{ij}$ characteristic.
     {The phase couplings $|J^{(\Phi)}_{ij}|$ (bottom row) show a more diffuse structure. For the GHZ state~(d), the relative phases between computational basis states are simple (a single phase between $|0\rangle^{\otimes N}$ and $|1\rangle^{\otimes N}$), so the phase RBM learns a homogeneous pattern. For the Bell-pair state~(e), the network shows weaker but visible two-qubit modularity. For the heterogeneous state~(f), both $|J^{(A)}|$ and $|J^{(\Phi)}|$ show complex patterns reflecting the non-trivial local basis transformations. Overall, $|J^{(A)}_{ij}|$ captures the dominant outcome probabilities, whereas $|J^{(\Phi)}_{ij}|$ controls the relative phases.}
  }
  \label{fig:J-maps}
\end{figure}

\subsubsection{Effective qubits couplings}
\label{subsec:interpret-J}

{

We first relate the RBM representation of Eq.~\eqref{eq:rbm-X-generic} to
an effective Ising--like model on the visible spins \cite{Rrapaj2021,Decelle2024,Decelle2025}.  For definiteness we
consider the amplitude RBM and suppress the superscript \((A)\) on the
weights. Writing
\begin{equation}
  A_\theta(\mathbf{s})
  =
  \sum_i a_i s_i
  + \sum_h
      \log\!\left[
        2\cosh\!\Bigl(
          b_h + \sum_i W_{ih} s_i
        \Bigr)
      \right],
  \label{eq:A-RBM-interpret}
\end{equation}
 {Here we specialize to the amplitude network, using the notation $(a_i, b_h)$ for the visible and hidden biases in place of the generic labels $(x_i, y_h)$ of Eq.~\eqref{eq:rbm-X-generic}, while $s_i\in\{\pm 1\}$ remains the spin configuration.}
We introduce \(x_h(\mathbf{s}) = b_h + \sum_i W_{ih} s_i\). For moderate
values of \(x_h\) one may expand \(\log(2\cosh x)\) around \(x=0\),
\begin{equation}
  \log\!\bigl(2\cosh x\bigr)
  = \log 2 + \frac{x^2}{2} - \frac{x^4}{12}
    + \mathcal{O}(x^6),
\end{equation}
which inserted into Eq.~\eqref{eq:A-RBM-interpret} yields a quadratic
approximation
\begin{equation}
  A_\theta(\mathbf{s})
  \approx
  A_0
  + \sum_i h_i s_i
  + \frac{1}{2}\sum_{i,j} J_{ij} s_i s_j,
  \label{eq:A-quadratic}
\end{equation}
where \(A_0\) is a constant. Using
\(x_h^2(\mathbf{s}) = b_h^2 + 2b_h\sum_i W_{ih}s_i + \sum_{i,j}W_{ih}W_{jh}s_is_j\)
one finds
\begin{equation}
  J_{ij}
  = \sum_h W_{ih} W_{jh}
     + \text{higher-order terms}.
  \label{eq:J-ij-from-RBM}
\end{equation}
An analogous construction applied to the phase RBM yields a matrix
\(J^{(\Phi)}\).
 
}

Figure~\ref{fig:J-maps} presents
\(|J^{(X)}|\), ($X\in \{A, \Phi\}$; amplitude: top row, phase: bottom row). The coupling element is non-vanishing when qubits \(i\) and \(j\) couple strongly to
overlapping sets of hidden units, and small when their hidden--layer
connectivity is largely disjoint. The visible structures
in \(|J^{(A)}|\) align with the basis-aggregated two-body correlations, Eq.~\eqref{eq:Cij-def}, and negativity, Eq.~\eqref{eq:negativity_eigs}, see Fig.~\ref{fig:Cij-maps}. In particular, the Bell--pair state
\(\ket{\psi_{\mathrm{B}}}\) produces three isolated \(2\times 2\) blocks
in \(|J^{(A)}|\), mirroring the three strong links in \(C_{ij}\). The GHZ
state \(\ket{\psi_{\mathrm{A}}}\) leads to a more homogeneous pattern,
consistent with the absence of a small--cluster decomposition at the
two--body level. For the heterogeneous state \(\ket{\psi_{\mathrm{C}}}\),
written in locally rotated bases, the weight--space structure of amplitude RBM, $|J^{(A)}_{ij}|$, reveals structure captured by negativity, see Fig.~\ref{fig:Cij-maps}(f), while the blocks of GHZ states are not visible in amplitude RBM, reflecting the increased complexity of its
representation in the computational basis. The effectiveness of this diagnostic depends on how simply the target
state is represented in the computational basis, since
\(A_\theta(\mathbf{s})\) and \(\Phi_\theta(\mathbf{s})\) are defined in
that basis.  

\begin{figure}[t]
  \centering
  \includegraphics[width=0.99\linewidth]{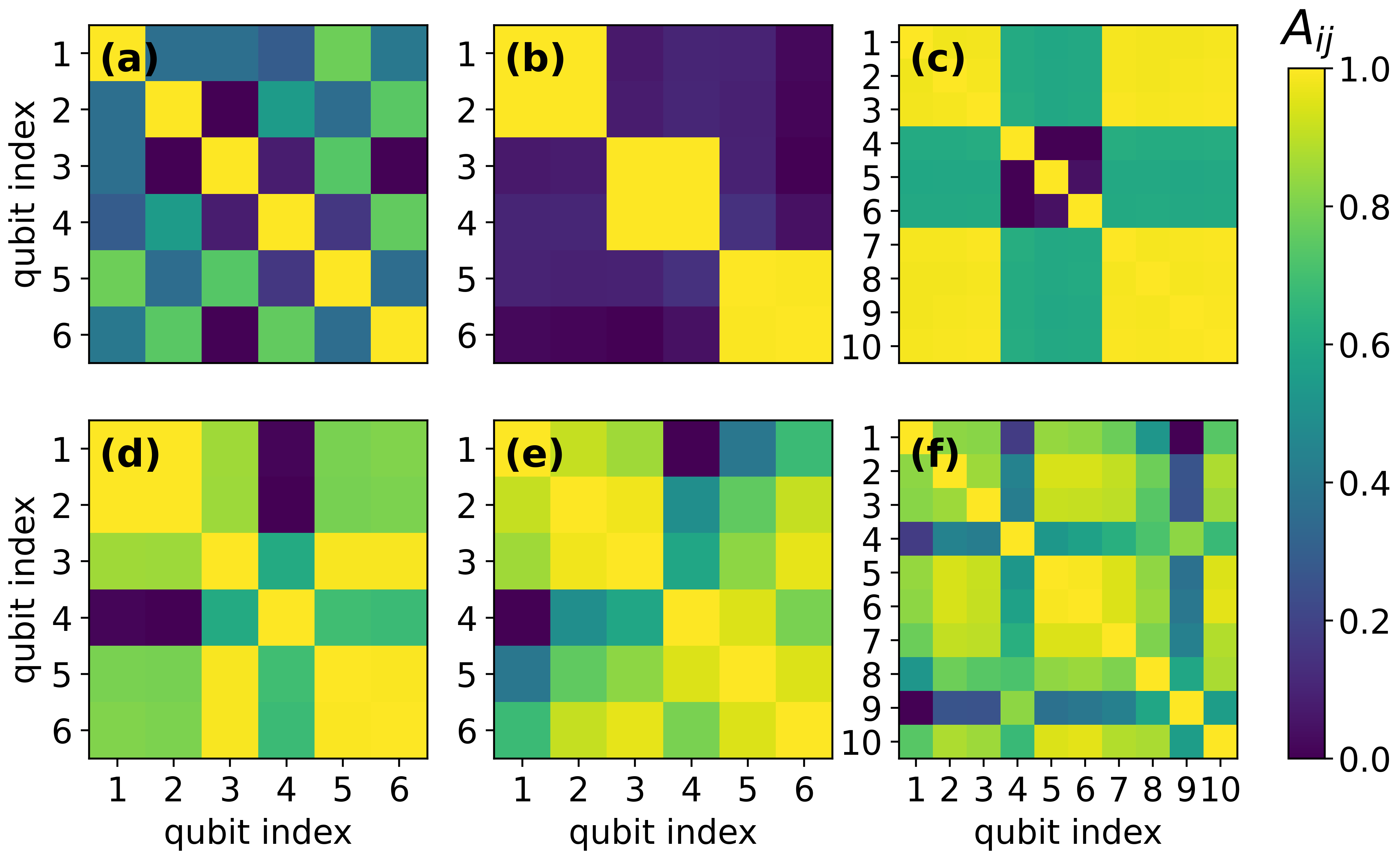}
  \caption{%
    Row-based affinity matrices for the unrestricted NQS. Each panel shows
    the normalized affinity \(A^{(X)}_{ij}\) defined in
    Eq.~\eqref{eq:affinity-def}, with \(X=A\) (amplitude, top row) or
    \(X=\Phi\) (phase, bottom row), computed from the row vectors of the
    weight matrices \(W^{(X)}\) of the trained unconstrained NQS. Columns
    correspond to the three pure target states \(\ket{\psi_{\mathrm{A}}}\),
    \(\ket{\psi_{\mathrm{B}}}\), and \(\ket{\psi_{\mathrm{C}}}\) as in
    Fig.~\ref{fig:J-maps}. Large values of \(A^{(X)}_{ij}\) indicate that
    qubits \(i\) and \(j\) couple to the hidden layer in a similar way.
    For the product of three Bell pairs \(\ket{\psi_{\mathrm{B}}}\) the
    amplitude affinity exhibits three nearly independent \(2\times 2\)
    blocks, while for the ten-qubit state \(\ket{\psi_{\mathrm{C}}}\) it
    reveals three clusters of sizes \(3\), \(3\), and \(4\), consistent
    with the intended tensor--product structure. The GHZ state
    \(\ket{\psi_{\mathrm{A}}}\) does not exhibit a pronounced block-diagonal pattern, consistent with the absence of a natural small-cluster decomposition.%
  }
  \label{fig:affinity-maps}
\end{figure}

\subsubsection{Row--based affinity between visible units}
\label{subsec:interpret-affinity}

Whereas \(J^{(X)}\) emphasizes pairwise interaction strength mediated by
shared hidden units, a complementary question is whether different qubits
play similar roles within the RBM. This is captured by comparing the
rows of \(W^{(X)}\). Let
\begin{equation}
  \mathbf{v}_i^{(X)}
  =
  \bigl(
    W^{(X)}_{i1},\ldots,W^{(X)}_{iM}
  \bigr)\in\mathbb{R}^{M}
\end{equation}
denote the row vector associated with qubit \(i\). We define squared
row distances
\begin{equation}
  d_{ij}^{2}
  =
  \bigl\|\mathbf{v}_i^{(X)}-\mathbf{v}_j^{(X)}\bigr\|_{2}^{2},
  \label{eq:row-distance}
\end{equation}
and rescale them into a normalized affinity
\begin{equation}
  A^{(X)}_{ij}
  =
  1-\frac{d_{ij}^{2}}{\max_{k,\ell} d_{k\ell}^{2}},
  \qquad
  A^{(X)}_{ii}=1,
  \label{eq:affinity-def}
\end{equation}
with \(X\in\{A,\Phi\}\). Large \(A^{(X)}_{ij}\) indicates that qubits
\(i\) and \(j\) couple to the hidden layer in a similar manner.

Figure~\ref{fig:affinity-maps} shows that \(A^{(X)}\) provides a clear
clustering signal that is consistent with the correlation 
\(C_{ij}\) and the coupling matrices \(J^{(X)}\). For the Bell--pair state
\(\ket{\psi_{\mathrm{B}}}\) the affinity reveals three nearly independent
\(2\times2\) blocks, while for the heterogeneous state
\(\ket{\psi_{\mathrm{C}}}\) it groups qubits into three clusters of sizes
\(3\), \(3\), and \(4\), recovering the intended tensor--product
structure directly at the level of learned parameters. The
\(A^{(X)}\) serves as a data--driven proxy for subsystem structure and
can be used to propose candidate partitions \(\mathcal{P}\) for the SNQS
hierarchy.

The metrics considered here, \(C_{ij}\), \(J^{(X)}\), and
\(A^{(X)}\), 
provide a compact
tomography--free representation of pairwise correlations that can be
extracted from raw measurement bitstrings and is faithfully reproduced by
the trained unrestricted NQS. These provide a principled route from an unrestricted fit
to candidate separability partitions for structured NQS models.

\section{Conclusions}
\label{sec:conclusion}

We have recast the problem of certifying non-$k$-separability and
entanglement depth as an experimentally driven, likelihood-based
model-comparison task between neural quantum states.  Instead of
reconstructing a density matrix and subsequently applying entanglement
criteria, we compare the optimized negative log-likelihoods of a fully
entangling neural ansatz and a hierarchy of partition-constrained
structured neural quantum states.  Within this hierarchy, each
multipartition $\mathcal{P}$ defines a  {$l$}-producible model family whose
maximal block size bounds the entanglement depth, and robust positive
likelihood gaps provide an operational witness that the data are
incompatible with that separability class.

A key ingredient is the use of structured neural quantum states for both
pure and mixed targets.  For pure states, separability is built into the
architecture via masked connectivity that enforces a given product
structure.  For mixed states, we promote these structured states to
low-rank ensembles so that classical correlations between blocks can be
captured without generating entanglement across the cuts.  In this way
the same likelihood-gap criterion applies uniformly in the presence or
absence of noise and distinguishes classical noise-induced correlations
from genuinely quantum ones.

Beyond certification, we considered an interpretability scheme for neural
quantum states that extracts structural information about the underlying
many-body state directly from the trained NQS parameters.  Rather than
treating the unrestricted ansatz as an uninterpretable one, we show that its learned
weights encode coarse entanglement features such as natural qubit
groupings and cluster structure.  This provides qualitative insight into
the organization of multipartite entanglement directly from measurement
bitstrings, without explicit state reconstruction, and can be used to
guide the design of structured ansatz families for further hypothesis
testing.

Our numerical experiments indicate that useful bounds on non-$k$-separability, together with qualitative information on the entanglement
structure, can be extracted from finite-shot local Pauli data and
relatively modest neural architectures, without full tomography or an
exhaustive scan over partitions.
 {In unconstrained scenarios, the interpretability diagnostics of Sec.~\ref{sec:interpretability} (effective couplings $J^{(X)}_{ij}$ and affinities $A^{(X)}_{ij}$) can guide partition selection directly from a single unrestricted training run, reducing the search overhead.}
 {The dominant computational cost in the current implementation is the exact enumeration of $2^N$ configurations; for larger systems, sampling-based NLL estimators and more expressive autoregressive neural quantum states~\cite{Sharir2020} can replace exact enumeration while preserving the partition-constrained structure that underpins the certification hierarchy.}
This generative capability distinguishes our method from randomized measurement protocols~\cite{elben2023randomized}, such as classical shadows~\cite{Huang2020}, which have successfully addressed the scalability of estimating scalar observables and fidelities. While highly efficient for specific properties, shadow-based estimators do not yield a generative model of the system's statistics, nor do they directly expose the granular structure of multipartite correlations. In contrast, our results demonstrate that structured NQS provide a best-fit generative approximation of the state itself, allowing one to not only certify entanglement depth but also inspect the learned weights to recover physical insights---such as effective coupling graphs and qubit affinity---that remain accessible only through a model-based approach.

Several research directions remain open, including a systematic treatment of statistical uncertainty in likelihood gaps.
Beyond the systematic treatment of statistical uncertainty, the interpretable weight diagnostics introduced here pave the way for automated model selection. Future work could employ Graph Neural Networks (GNNs) to process the effective interaction graphs derived from the unconstrained NQS. By mapping these topological features to qubit clusterings, GNNs could serve as a data-driven proposal mechanism, automatically identifying the most relevant candidate partitions to test within the structured ansatz hierarchy \cite{bronstein2021geometric}.

\section*{Code availability}
The code used to generate the results and figures in this work is publicly available at
\url{https://github.com/MarcinPlodzien/Neural-Quantum-States-for-entanglement-depth-certification},
\cite{code_plodzien_neural_quantum_states_repo_2025}.

\section*{Acknowledgments}
We thank Grzegorz Rajchel-Mieldzioć, Arnau Riera, and Anna Dawid for useful comments.
We acknowledge RES resources provided by Barcelona Supercomputing Center in Marenostrum 5 to NNO-2025-3-0004, and funds from MICIU/AEI/10.13039/501100011033/ FEDER, UE.

\bibliography{references}

\appendix
\section{RBM parameterization of the NQS and SNNS ansatzes}
\label{app:numerical_details}

This Appendix records the precise RBM architectures used in the numerical experiment  \cite{code_plodzien_neural_quantum_states_repo_2025}. We focus on the concrete parameterization implemented
in the code (parameter tensors and their dimensions), and avoid re-deriving
expressions already introduced in the main text.

\subsection{Visible variables and notation}
The visible layer is an Ising encoding of computational-basis bitstrings,
$s\in\{-1,+1\}^N$, obtained from $x\in\{0,1\}^N$ by $s_i=2x_i-1$. All model
probabilities required for training are evaluated by explicit enumeration over
all $2^N$ configurations.

The pure-state NQS wave function is represented as a product of an amplitude
RBM and an independent phase RBM,
\begin{equation}
\psi_\theta(s)=\exp\!\big[\mathcal{A}_\theta(s)\big]\,
\exp\!\big[i\,2\pi\,\Phi_\theta(s)\big].
\end{equation}
Both $\mathcal{A}_\theta$ and $\Phi_\theta$ are computed from RBM-like hidden
layers with the visible variables marginalized analytically (i.e., in
``free-energy'' form). The amplitude RBM uses $H_{\rm amp}$ hidden units with
parameters
\begin{equation}
a\in\mathbb{R}^{N},\qquad
b\in\mathbb{R}^{H_{\rm amp}},\qquad
W\in\mathbb{R}^{N\times H_{\rm amp}}.
\end{equation}
The phase RBM uses $H_{\rm phase}$ hidden units with parameters
\begin{equation}
c\in\mathbb{R}^{N},\qquad
d\in\mathbb{R}^{H_{\rm phase}},\qquad
U\in\mathbb{R}^{N\times H_{\rm phase}}.
\end{equation}
In the implementation, the phase hidden contributions are passed through the
bounded nonlinearity
\begin{equation}
g(z)=\frac{2}{\pi}\arctan[\tanh(z)]+\frac{1}{2},
\end{equation}
and the overall complex phase factor is $\exp(i2\pi\,\Phi_\theta)$.
Throughout training and evaluation, the state vector $\psi_\theta$ is
normalized explicitly, $\psi_\theta\to \psi_\theta/\|\psi_\theta\|_2$.

The ensemble model represents a density matrix as a convex mixture of pure NQS
states,
\begin{equation}
\rho_\theta=\sum_{k=1}^{K} w_k\,|\psi_{\theta_k}\rangle\langle\psi_{\theta_k}|,
\qquad
w_k=\frac{e^{\ell_k}}{\sum_{j=1}^K e^{\ell_j}}.
\end{equation}
Each component $|\psi_{\theta_k}\rangle$ is parameterized by its own set of RBM
weights and biases $\theta_k=\{a^{(k)},b^{(k)},W^{(k)},c^{(k)},d^{(k)},U^{(k)}\}$,
with the same dimensions as in the pure-state ansatz. The mixing weights are
determined by unconstrained logits $\ell\in\mathbb{R}^K$.

For constrained NQS, the qubits are partitioned into disjoint clusters $\{C_b\}_b$ with
$N_b=|C_b|$ and $\sum_b N_b=N$. Each block is assigned an RBM-based NQS of the
same functional form as the global ansatz, but restricted to the visible spins
$s_{S_b}$. A pure constrained NQS wave function is the product of  wave functions,
$\psi_{\rm SQNS}(s)=\prod_b \psi^{(b)}_{\theta_b}(s_{C_b})$,
with independent parameters
$\theta_b=\{a_b,b_b,W_b,c_b,d_b,U_b\}$ of dimensions
$
a_b,c_b\in\mathbb{R}^{N_b},\quad
b_b\in\mathbb{R}^{H^{(b)}_{\rm amp}},\quad
d_b\in\mathbb{R}^{H^{(b)}_{\rm phase}},\quad
W_b\in\mathbb{R}^{N_b\times H^{(b)}_{\rm amp}},\quad
U_b\in\mathbb{R}^{N_b\times H^{(b)}_{\rm phase}}.
$
In the simulations, if block hidden sizes are not set explicitly, one uses
$H^{(b)}_{\rm amp}=H_{\rm amp}$ and $H^{(b)}_{\rm phase}=H_{\rm phase}$.
   
We fixed the RBM hidden-layer sizes to
$H_{\rm amp}=64$ and $H_{\rm phase}=64$, and initialize all trainable parameters
(including ensemble logits) from an i.i.d.\ normal distribution with standard
deviation $10^{-2}$. Calculations are performed in double precision. Optimization is carried out by minimizing the
negative log-likelihood with Adam optimiser using a cosine-decay learning-rate schedule
(from $\eta_0=5\times 10^{-3}$ to $10^{-2}\eta_0$ over $6000$ steps) implemented in JAX/Optax~\cite{deepmind2020jax}.

\end{document}